\documentclass[10pt,amssymb]{iopart}
\usepackage{iopams,epsf}
\usepackage{mathptm,times}

\catcode`\@ 11
\eqnobysec
\def\eqref#1{(\ref{#1})}

\def\journal#1&#2,#3{\begingroup \let\journal=\dummyjournal
               \it #1\unskip~\bf\ignorespaces #2\rm, #3 \endgroup}
\def\jour{\ifnum\count255<100 \advance\count255 by 1900 \fi\number\count255 }
\def\dummyjournal{\errmessage{Reference foul up: nested \journal macros}}

\newcounter{theorem}
\@addtoreset{theorem}{section}
\renewcommand\thetheorem{\arabic{section}.\arabic{theorem}}
\newenvironment{definition}{\par\medskip\noindent\begingroup{\bf Definition
             \stepcounter{theorem}\thetheorem.}\ \itshape
             \def\@currentlabel{\thetheorem}}{\endgroup\par\medskip}
\newenvironment{lemma}{\par\medskip\noindent\begingroup{\bf Lemma
             \stepcounter{theorem}\thetheorem.}\ \itshape
             \def\@currentlabel{\thetheorem}}{\endgroup\par\medskip}
\newenvironment{theorem}{\par\medskip\noindent\begingroup{\bf Theorem
             \stepcounter{theorem}\thetheorem.}\ \itshape
             \def\@currentlabel{\thetheorem}}{\endgroup\par\medskip}
\newenvironment{proposition}{\par\medskip\noindent\begingroup{\bf Proposition
             \stepcounter{theorem}\thetheorem.}\ \itshape
             \def\@currentlabel{\thetheorem}}{\endgroup\par\medskip}
\newenvironment{remark}{\par\medskip\noindent\begingroup{\bf Remark
             \stepcounter{theorem}\thetheorem.}\ 
             \def\@currentlabel{\thetheorem}}{\endgroup\par\medskip}
\newenvironment{corollary}{\par\medskip\noindent\begingroup{\bf Corollary
             \stepcounter{theorem}\thetheorem.}\ \itshape
             \def\@currentlabel{\thetheorem}}{\endgroup\par\medskip}
\newenvironment{proof}{\par\noindent{\bf Proof.} }{\proofbox\par\medskip}

\def\proofbox{\hfill{\ensuremath\Box}}

\newcommand\partialderiv[3][]{\frac{\partial^{#1}#2}{\partial {#3}^{#1}}}
\newcommand\deriv[3][]{\frac{d^{#1}#2}{d{#3}^{#1}}}
\let\truesum@=\sum
\def\sum{\mathop{\textstyle\truesum@}\limits}
\def\txtprod{\mathop{\textstyle\prod}\limits}
\def\txtfrac#1#2{{\textstyle\frac{#1}{#2}}}
\def\diag{\mathop{\rm diag}\nolimits}
\def\sech{\mathop{\rm sech}\nolimits}
\def\O#1{^{(#1)}}
\def\half{{\textstyle\frac12}}
\def\Nm{{\ensuremath{N_-}}}
\def\Np{{\ensuremath{N_+}}}
\begin{document}
\title[On a family of solutions of the KP equation satisfying the
Toda lattice hierarchy]%
{On a family of solutions of the KP equation
which also satisfy the Toda lattice hierarchy}
\author{Gino Biondini and Yuji Kodama
\unskip\footnote[3]{To whom correspondence should be addressed
(kodama@math.ohio-state.edu)}}
\address{Department of Mathematics, Ohio State University,\\
231 West 18th Ave, Columbus, OH 43210}

\begin{abstract}
We describe the interaction pattern in the $x$-$y$ plane for a family
of soliton solutions of the Kadomtsev-Petviashvili (KP) equation,
\[
(-4u_{t}+u_{xxx}+6uu_x)_{x}+3u_{yy}=0.
\]
Those solutions also satisfy the finite Toda lattice hierarchy.
We determine completely their asymptotic patterns for $y\to \pm\infty$,
and we show that all the solutions (except the one-soliton solution) are of
{\it resonant} type, consisting of arbitrary numbers of line solitons in
both aymptotics; that is, arbitrary $N_-$ incoming solitons for $y\to -\infty$
interact to form arbitrary $N_+$ outgoing solitons for $y\to\infty$.
We also discuss the interaction process of those solitons, and show that
the resonant interaction creates a {\it web-like} structure having
$(N_--1)(N_+-1)$ holes.
\end{abstract}


\submitto{\JPA}

%%%%%%%%%%%%%%%%%%%%%%%%%%%%%%%%%%%%%%%%%%%%%%%%%%%%%%%%%%%%%%%%%%%%%%%%%%%%%%%

\section{Introduction}

In this paper we study a family of solutions of the
Kadomtsev-Petviashvili (KP) equation
\begin{equation}
\frac{\partial}{\partial x}\left(-4\frac{\partial u}{\partial t}
+\frac{\partial^3u}{\partial x^3}+6u\frac{\partial u}{\partial x}\right)
            + 3 \frac{\partial ^2u}{\partial y^2}=0\,
\label{e:kp}
\nonumber
\end{equation}
which can be written in the bilinear form~\cite{hirota1976},
\begin{equation}
\big[-4D_xD_t+D^4_x+3D^2_y\,\big]\,\tau\cdot\tau=0\,.
\label{e:bilinearKP}
%\nonumber
\end{equation}
Here $D_x$, $D_y$ and $D_t$ are the Hirota derivatives, e.g.,
$D_x^m f\cdot g= (\partial_x-\partial_{x'})^m f(x,y,t)g(x',y,t)|_{x=x'}$
etc., and $u$ is obtained from the tau-function $\tau(x,y,t)$ as
\begin{equation}
u(x,y,t)= 2\partialderiv[2]{}x\log \tau(x,y,t)\,.
\label{e:kpsoln}
\end{equation}

It is well-known that some solutions of the KP equation can be obtained by
the Wronskian form $\tau=\tau_M$ (see Appendix and also~\cite{nimmo1983}),
with
\begin{equation}
\tau_M=\mathop{\rm Wr}\,(f_1,\dots,f_M):=
\left|\begin{array}{ccc}
f_1\O0 &\cdots &f_M\O0\\
\vdots &\ddots &\vdots\\
f_1\O{M-1} &\cdots &f_M\O{M-1}
\end{array}\right|,
\label{e:tau}
\end{equation}
where $f_i\O{n}= \partial^n f_i/\partial x^n$, and
$\{f_i(x,y,t)\,|~i=1,\dots,M\}$ is a
linearly independent set of $M$ solutions of the equations,
\begin{equation}
\partialderiv {f_i}y= \partialderiv[2]{f_i}x\,, \qquad
\partialderiv {f_i}t= \partialderiv[3]{f_i}x\,, \qquad
\mathrm{for}~\quad 1\le i\le M.
%\label{e:linear}
\nonumber
\end{equation}
For example, the two-soliton solution of the KP equation is obtained by the set
$\{f_1,f_2\}$, with
\begin{equation}
f_i= e^{\theta_{2i-1}}+e^{\theta_{2i}}\,, \qquad i=1,2\,,
\label{e:f2solitons}
%\nonumber
\end{equation}
where the phases $\theta_j$ are given by linear functions of $(x,y,t)$,
\begin{equation}
\theta_j(x,y,t)= -k_jx + k_j^2y-k_j^3t+ \theta_j^0\,, \qquad j=1,\dots,4\,,
\label{e:theta}
\end{equation}
with $k_1<k_2<k_3<k_4$. This ordering is sufficient for the solution~$u$
to be nonsingular.  (The ordering $k_1\ne k_2<k_3\ne k_4$ is needed
for the positivity of~$\tau_2$.)\, Note, for example, that
if $k_1<k_3<k_2<k_4$,
$\tau_2$ takes zero and the solution blows up at some points in~$(x,y,t)$.
The formula~\eqref{e:f2solitons} can be extended to the
$M$-soliton solution with $\{f_1,\cdots,f_M\}$~\cite{hirota1988}.

On the other hand,
it is also known that the solutions of the finite Toda lattice hierarchy are
obtained by the set of tau-functions
$\{\tau_M\,|~M=1,\dots,N\}$ with the choice of $f$-functions,
\begin{equation}
\label{e:ftoda}
\left\{
\begin{array}{ll}
\\[-3ex]
f_1= \sum_{i=1}^N e^{\theta_i}=:f\,,\\[2ex]
f_i= f\O{i-1}\,,\qquad 1< i\le M\le N\,,\\[0.4ex]
\end{array}
\right.
\end{equation}
where the phases $\theta_i$, $1\le i\le N$ are given in the
form~\eqref{e:theta} (see for example \cite{nakamura1995}).
This implies that each tau-function $\tau_M$ gives a solution of the
KP equation.
If the $f$-functions are chosen according to~\eqref{e:ftoda}, the
tau-functions are then given by the Hankel determinants
\begin{equation}
\tau_M=
\left|\begin{array}{ccc}
f\O0 &\cdots &f\O{M-1}\\
\vdots &\ddots &\vdots\\
f\O{M-1} &\cdots &f\O{2M-2}
\end{array}\right|,
\qquad {\rm for}\quad 1\le M\le N\,.
\label{e:todatau}
\end{equation}
Note here that $\tau_N=\,C\,\exp(\theta_1+\cdots+\theta_N)$,
with $C={}$constant, yielding the trivial solution, and $\tau_M$ and
$\tau_{N-M}$
produce the same solution with the symmetry $(x,y,t)\to (-x,-y,-t)$,
due to the duality of the determinants (i.e., the duality of the
Grassmannians Gr($M,N$) and Gr($N-M,N$); see also Lemma~\ref{l:tauN}).
The finite Toda lattice hierarchy is defined in the Lax
form~\cite{flaschka1974}
\begin{equation}
\partialderiv L{t_n}=[B_n,L]\,,\qquad n=1,\dots,N-1\,,
%\label{e:LaxToda}
\nonumber
\end{equation}
%where $t_1=x$, $t_2=y$, $t_3=t$, and
where the Lax pairs $(L,B_n)$ are given by
\begin{eqnarray*}
L= \left(\begin{array}{ccccc}
b_1 &a_1 &0 &\cdots &0\\
a_1 &b_2 &a_1 &\ddots &\vdots\\
0 & a_2 &\ddots &\ddots &0\\
\vdots  &\ddots &\ddots &b_{N-1} &a_{N-1}\\
0 &\cdots &0 &a_{N-1} &b_N
\end{array}\right)\,,
\\[1ex]
B_n=\half\left((L^n)_{>0}-(L^n)_{<0}\right)\,\,,
\end{eqnarray*}
and where $C_{>0}$ ($C_{<0}$) denotes the strictly upper (lower)
triangular part of a matrix~$C$.
Here the flow parameters $t_i$'s are chosen as $t_1=x,~t_2=y$ and
$t_3=t$ for the KP equation.
The functions $(a_i,b_j)$ are expressed by
\begin{equation}
\label{e:abToda}
\left\{
\begin{array}{ll}
\displaystyle{a^2_n= \frac{\tau_{n+1}\tau_{n-1}}{\tau_n^2}\,,\qquad
n=1,\dots,N-1\,,}\\
\displaystyle{b_n= \deriv{}t\log\frac{\tau_n}{\tau_{n-1}}\,,\qquad
n=1,\dots,N\,,}
\end{array}\right.
\end{equation}
where $\tau_0=1$.
Then the tau-functions $\tau_n$ satisfy the bilinear equations
\begin{equation}
\displaystyle{\frac{1}{2} D_x^2\,\tau_n\cdot\tau_n
             = \tau_n\tau_{n,xx}-(\tau_{n,x})^2
             = \tau_{n+1}\tau_{n-1}\,,}
%\label{e:bilinear}
\nonumber
\end{equation}
which are just the Jacobi formulae for the determinants $D:=\tau_{n+1}$, i.e.
\begingroup
\def\D[#1,#2]{D\bigg[\!\!\begin{array}{cc}#1\\#2\end{array}\!\!\bigg]}
\begin{equation}
\D[n+1,n+1]\D[n,n]-\D[n,n+1]\D[n+1,n]= \D[{n,\,n+1},{n,\,n+1}]\,D\,\,.
\label{jacobi}
\end{equation}
Here $\D[{i,\,j},{k,\,l}]$ denotes the determinant obtained by deleting the
$i$-th and $j$-th rows and the $k$-th and the $l$-th column in $D$
\cite{gantmacher1959}.
\endgroup

\begin{remark}
According to the Sato theory (see for example~\cite{ohta1988}), these
bilinear equations for the KP equation and
the Toda lattice hierarchy are the Pl\"ucker relations
with proper definitions of the Pl\"ucker coordinates $\tau_Y$ labeled by
Young diagrams $Y=(\ell_1,\ell_2)$, with $\ell_1\le \ell_2$ giving the
numbers of boxes in $Y$,
\begin{equation}
\tau_{(0,0)}\tau_{(2,2)}-\tau_{(0,1)}\tau_{(1,2)}
             +\tau_{(0,2)}\tau_{(1,1)}=0\,.
             \label{plucker}
\end{equation}
For the KP equation, those Pl\"ucker coordinates are
related to the derivatives of the tau-function $\tau_M$,
\[
\left\{
\begin{array}{llll}
\displaystyle\tau_{(0,0)}=\tau_M,\\[0.8ex]
\displaystyle\tau_{(0,1)}=\partial_x \tau_M,\\[0.8ex]
\displaystyle\tau_{(0,2)}=\txtfrac12(\partial_x^2+\partial_y)\tau_M,\\[0.8ex]
\displaystyle\tau_{(1,1)}=\txtfrac12(\partial_x^2-\partial_y)\tau_M\\[0.8ex]
\displaystyle\tau_{(1,2)}=\txtfrac13(\partial_x^3-\partial_t)\tau_M,\\[0.8ex]
\displaystyle\tau_{(2,2)}=\txtfrac1{12}(\partial_x^4-4\partial_x\partial_t
   +3\partial_y^2)\tau_M\,.
\end{array}
\right.
\]
Then the Hirota bilinear equation~\eqref{e:bilinearKP} is equivalent
to the Pl\"ucker relation~\eqref{plucker}. For the Toda lattice equation,
the Jacobi formula~\eqref{jacobi} can be considered as~\eqref{plucker}
with the identification $\tau_{(0,0)}=D$ etc.
\end{remark}

We should also recall that the solutions of the Toda lattice equation
show the sorting property of the Lax matrix $L$~\cite{moser1975}; that is,
\begin{equation}
L ~\longrightarrow~\left\{\begin{array}{ll}
\diag(\lambda_1,\dots,\lambda_N) &\quad{\rm as}~~x\to\infty\,,\\
\diag(\lambda_N,\dots,\lambda_1) &\quad{\rm as}~~x\to-\infty\,,
\end{array}\right.
\label{e:todasorting}
\end{equation}
where $\lambda_1>\lambda_2>\cdots>\lambda_N$ are the eigenvalues of $L$.
These eigenvalues are related to the parameters $k_i$ in~\eqref{e:theta}
as $\lambda_i=-k_i$ (see below).
%In the present paper, we will describe the evolution of $\diag(L)$.

In this paper, we are concerned with the behavior of the
KP solutions~\eqref{e:kpsoln} whose tau-functions are
given by~\eqref{e:todatau}.
We describe the patterns of the solutions in the $x$-$y$ plane
where each soliton solution of the KP equation is asymptotically
expressed as a line, namely,
\[
x= c_\pm y + \xi_\pm\qquad\mathrm{for}\quad y\to\pm\infty
\]
with appropriate constants $c_\pm$ and $\xi_\pm$ for a fixed~$t$.
In particular, we found that all the solutions
(except the one-soliton solution) are ``resonant'' solitons
in the sense that these solutions are different from ordinary
multi-soliton solutions.
The difference appears in the process of interaction, which results,
for example, in a different number of solitons (or lines) asymptotically
as $y\to\infty$ or $y\to-\infty$.

In our main result (Theorem \ref{t:asymptotic}) we show that
for the solution with the tau-function given by~\eqref{e:todatau}
with~\eqref{e:ftoda},
the number of solitons in asymptotic stages as $y\to\pm\infty$,
denoted by $N_+$ and $N_-$, is given by
\[
N_+=M\,,\qquad N_-=N-M\,.
\]
Thus, the total number~$N$ of exponential terms in the function $f$
in~\eqref{e:ftoda} gives the total number of solitons present in both
asymptotic limits, i.e., $N=\Nm+\Np$, and the number of outgoing
solitons $\Np$ is given by the size of the Hankel determinant
\eqref{e:todatau}.
We call these solutions ``(\Nm,\Np)-solitons''.
In particular, if $N=2\Np=2\Nm$, the solution describes an
$\Np$-soliton having the same set of line solitons in each
asymptotics for $y\to\pm\infty$.
However, these multi-soliton solutions also differ from the ordinary
multi-soliton solutions of the KP equation.
The ordinary $n$-soliton solution of the KP equation is described by
$n$~intersecting line solitons with a phase shift at each interaction
point.
If we ignore the phase shifts, these $n$ lines form $(n-1)(n-2)/2$
bounded regions in the generic situation.
However, the number of bounded regions for the (resonant)
$\Np$-soliton solution with~\eqref{e:todatau} is found
to be~$(\Np-1)^2$;
for example, even in the case of a two-soliton solution there is
one bounded region as a result of the resonant interaction.
In general, we show in Proposition~\ref{p:holes} that for the
case of a~$(N_-,N_+)$-soliton solution,
the number of bounded regions (holes) in the graph is given by
$(N_--1)(N_+-1)$, except at finite values of $t$ in the temporal evolution.

These resonant $N_+$-soliton solutions are similar to some of the
solitons of the coupled KP (cKP)
hierarchy recently studied in Ref.~\cite{JPhysA2002v35p6893},
where such solutions were called ``spider-web-like'' solutions.
The analysis of finding web~structure that we describe in the
present study may also be applied to the case of the cKP hierarchy.

%%%%%%%%%%%%%%%%%%%%%%%%%%%%%%%%%%%%%%%%%%%%%%%%%%%%%%%%%%%%%%%%%%%%%%%%%%%%%%%

\section{Asymptotic analysis of the solutions}

Before we discuss the general case for the tau-function~\eqref{e:todatau}
with~\eqref{e:ftoda}, we present some simple cases corresponding to
a (1,1)-soliton and a (2,1)-soliton solution; the latter turns out to be
the resonant case of an ordinary 2-soliton solution of the KP equation.

As explained in the Appendix, we first note that
the $(\Nm,1)$-soliton can be described as the solution
of the Burgers equation~\eqref{e:burgers},
\[
\frac{\partial w_1}{\partial y}+2w_1\frac{\partial w_1}{\partial x}=
\frac{\partial^2w_1}{\partial x^2}, \quad {\rm with}\quad
w_1= - \partialderiv{ }x \log\tau_1\,.
\]
An explicit solution of this equation is a shock, which corresponds to the case
of $N=2=1+1$, i.e., $\tau_1=e^{\theta_1}+e^{\theta_2}$.
The solution $w_1$ is then given by
\[\begin{array}{lllll}
w_1&=&{\frac{1}{2}(k_1+k_2)+\frac{1}{2}(k_1-k_2){\rm
tanh}\frac{1}{2}(\theta_1-\theta_2) }\\[1.2ex]
&\longrightarrow&
\left\{
\begin{array}{ll}
k_1 &\quad{\rm as}~~x\to\infty,\\
k_2 &\quad{\rm as}~~x\to -\infty,
\end{array}
\right. \quad({\rm for}~~k_1<k_2)
\end{array}
\]
which leads to the one-soliton solution of the KP equation,
\begin{equation}
\label{e:1-soliton}
u= 2\partialderiv[2]{ }x \log\tau_1=
             \half(k_1-k_2)^2\sech^2\half(\theta_1-\theta_2)\,.
\end{equation}
In the $x$-$y$ plane, this solution describes a plane wave
$u=\Phi(k_xx+k_yy-\omega\,t)$ having the wavenumber vector
${\bf k}=(k_x,k_y)$ and
the frequency $\omega$,
\[
{\bf k}=(-k_1+k_2,~k_1^2-k_2^2)=:{\bf k}_{1,2},\quad
\omega=k_1^3-k_2^3=:\omega_{1,2}.
\]
Here $({\bf k},~\omega)$ satisfies the dispersion relation,
$4\omega k_x+k_x^4+3k_y^2=0$. We refer to the one-soliton solution
(\ref{e:1-soliton}) as a line soliton, which can be expressed by
a (contour) line, $\theta_1=\theta_2$, in the $x$-$y$ plane.
In this paper, since we discuss the pattern of soliton solutions
in the $x$-$y$ plane, we refer to $c=dx/dy$ as the velocity of the
line soliton in the $x$ direction; that is, $c=0$ indicates the
direction of the positive $y$-axis.

\begin{figure}[t!]
\medskip
\centerline{\epsfxsize0.475\textwidth\epsfbox{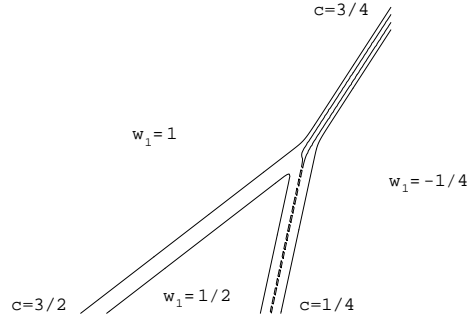}}
\caption{The confluence of two shocks of the Burgers equation
with $(k_1,k_2,k_3)=(-\frac14,\frac12,1)$, which also represents a
resonant soliton solution of the KP equation.
Here and in the following, unless indicated otherwise,
the horizontal and vertical axes are respectively $x$~and~$y$,
and the graph shows contour lines of the function
$u(x,y,t)=-2\partial_x\,w_1(x,y,t)$.}
\label{f:burgers}
\end{figure}

Now we consider the case of a (2,1)-soliton, whose tau-function is
given by
\[
\tau_1= e^{\theta_1} + e^{\theta_2} + e^{\theta_3}\,.
\]
This situation is explained in Ref.~\cite{whitham1974},
and the solution describes the confluence of two shocks.
Taking $k_1<k_2<k_3$ without loss of generality,
for $y\to-\infty$ the two shocks
(which correspond to line solitons for $u$)
have velocities $c_{1,2}=k_1+k_2$ and
$c_{2,3}=k_2+k_3$, and the single shock for $y\to\infty$
has velocity $c_{1,3}=k_1+k_3$.
This case is illustrated~figure~\ref{f:burgers} with
$(k_1,k_2,k_3)=(-\frac14,\frac12,1)$. A simple analysis (see below for more
details) shows that
the function $w_1=-\partial_x\log \tau_1$ takes the following asymptotic
values: $w_1\sim k_1=-\frac14$ for large $x$, $w_1\sim k_3=1$ for
large $-x$, and in the middle region for large $-y$, $w_1\sim k_2=\frac12$.

This Y-shape interaction represents a resonance of three line solitons.
The resonance conditions for three solitons with the wavenumber vectors
$\{{\bf k}_{i,j}\,|~1\le i<j\le 3\}$ and the frequencies
$\{\omega_{i,j}\,|~1\le i<j\le 3\}$
are given by
\begin{equation}
\label{resonant}
{\bf k}_{1,2}+{\bf k}_{2,3}={\bf k}_{1,3},\quad {\rm and}\quad
\omega_{1,2}+\omega_{2,3}=\omega_{1,3},
\end{equation}
which are trivially satisfied by those line solitons.
Here we point out that this solution is also the resonant case
of the ordinary 2-soliton solution of the KP equation.
As we mentioned earlier, the ordinary 2-soliton solution is given by the
$M=2$ tau-function~\eqref{e:tau} with~\eqref{e:f2solitons}.
The explicit form of the $\tau_2$-function is
\[
\tau_2= (k_1-k_3)\,e^{\theta_1+\theta_3}
    + (k_1-k_4)\,e^{\theta_1+\theta_4}
    + (k_2-k_3)\,e^{\theta_2+\theta_3}
    + (k_2-k_4)\,e^{\theta_2+\theta_4},
\]
where, as before, $\theta_i= -k_ix + k_i^2y - k_i^3t+ \theta_i^0$.
Note that if $k_2=k_3$, the $\tau_2$-function can be written as
\[
\tau_2= e^{\theta_1+\theta_2+\theta_4}\big[
             (k_1-k_3)\,\Delta\,e^{-\theta_4} + (k_1-k_4)\,e^{-\theta_2}
                + (k_2-k_4)\,e^{-\theta_1}\big]\,,
\]
where $\Delta=\exp(\theta_3^0-\theta_2^0)=$\,constant.
Since the exponential factor $ e^{\theta_1+\theta_2+\theta_4} $ gives
zero contribution to the solution~$u=2\partial_x^2\log\tau_2$,
the $\tau_2$-function is equivalent to the case of a
(2,1)-soliton solution (except the signs of the phases, and 
more precisely it is a (1,2)-soliton); that is, the resonant 
solution with confluence of solitons.
Note also that the condition $k_2=k_3$ is nothing else but the resonant
condition in Ref.~\cite{JFM1977v79p171},
and it describes the limiting case of an infinite phase shift
in the ordinary 2-soliton solution,
where the phase shift between the solitons as $y\to\pm\infty$ is
given by
\[
\delta= \frac{(k_1-k_3)(k_2-k_4)}{(k_2-k_3)(k_1-k_4)}\,.
\]
The resonant process for the $(\Nm,1)$-soliton
solutions of the KP equation can be expressed as a generalization
of the confluence of shocks discussed earlier.  This case
has been discussed in Ref.~\cite{LMP2002v62p91}.

We now discuss the general case of $(\Nm,\Np)$-soliton solutions.
In order to describe the asymptotic pattern of the solution associated
with the tau-function~\eqref{e:todatau},
we start with the following:

\begin{lemma}
\label{l:tauN}
Let $f$ be given by
\[
f=\sum_{i=1}^N e^{\theta_i}\,,\qquad\mathrm{with}\quad
\theta_i= -k_ix+k_i^2y+\theta_i^0\,.
\]
Then for $N=\Np+\Nm$ and $1\le \Np\le N-1$, the
tau-function defined by the Hankel determinant~\eqref{e:todatau}
has the form
\begin{equation}
\label{e:tauplus}
\displaystyle{\tau_\Np=
             \sum_{1\le i_1<\cdots<i_\Np\le N}
    \Delta(i_1,\dots,i_\Np)\,\,\exp\bigg(\sum_{j=1}^\Np\theta_{i_j}\bigg)\,,}
\end{equation}
where $\Delta(i_1,\dots,i_N)$ is the square of the van der Monde determinant,
\[
\Delta(i_1,\dots,i_N)=\txtprod_{1\le j<l\le \Np} (k_{i_j}-k_{i_l})^2\,.
\]
\end{lemma}

\begin{proof}
Apply the Binet-Cauchy theorem \cite{gantmacher1959} for
\[\fl
\tau_\Np= \det\left[
             \left(\begin{array}{cccc}
                e^{\theta_1} &e^{\theta_2} &\cdots &e^{\theta_N}\\
                k_1e^{\theta_1} &k_2e^{\theta_2} &\cdots &k_N e^{\theta_N}\\
                \vdots &\vdots &\ddots &\vdots\\
                k_1^{\Np-1}e^{\theta_1} &k_2^{\Np-1}e^{\theta_2} &\cdots
&k_N^{\Np-1}e^{\theta_N}
              \end{array}\right)
             \left(\begin{array}{ccccc}
                1 &k_1 &\cdots &\cdots &k_N^{\Np-1}\\
                1 &k_2 &\cdots &\cdots &k_N^{\Np-1}\\
                \vdots &\vdots &\ddots &\ddots &\vdots\\
                \vdots &\vdots &\ddots &\ddots &\vdots\\
                1 &k_N &\cdots &\cdots &k_N^{\Np-1}
              \end{array}\right)
             \right]\,.\
\]
\end{proof}

One should note from~\eqref{e:tauplus}
that the $\tau_{\Np}$-function contains {\it all} possible combinations of
$\Np$ phases from the set $\{\theta_j\,|~j=1,\dots,N\}$,
unlike the case of ordinary
multi-soliton solutions of the KP equation. For example, the
$\tau_2$-function for the
2-soliton solution with~\eqref{e:f2solitons} include only four terms, 
and is missing the combinations 
$\theta_1+\theta_2$ and $\theta_3+\theta_4$.  
This makes a crucial difference
on the interaction patterns of soliton solutions, as explained in this paper.
In particular, we will see that the $(\Nm,\Np)$-solitons are all of
resonant type in the sense that local structure of each interaction
point in those solitons
consists of either $(2,1)$- or $(1,2)$-solitons.

\begin{remark}
The $\tau_{\Np}$-function given by (\ref{e:todatau}) is positive definite,
and therefore the solution $u$ has no singularity. In general,
the Wronskian~\eqref{e:tau} takes zeros at some points
in the flow parameters.
The set of those points is called Painlev\'e divisor, whose geometry
has an interesting structure related to the Birkhoff stratification
of the Grassmannian~\cite{adler1994}.
Also, if one includes some exponential terms with negative coefficients
in~\eqref{e:ftoda}, the $\tau_M$-functions vanish on a set of
points in $(t_1,t_2,\cdots,t_{N-1})$. Then the set of those points
can be described as intersections of the Bruhat cells of the
flag manifold (see for example~\cite{casian2002}).
\end{remark}

\begin{figure}[t!]
\medskip
\centerline{\epsfxsize0.675\textwidth\epsfbox{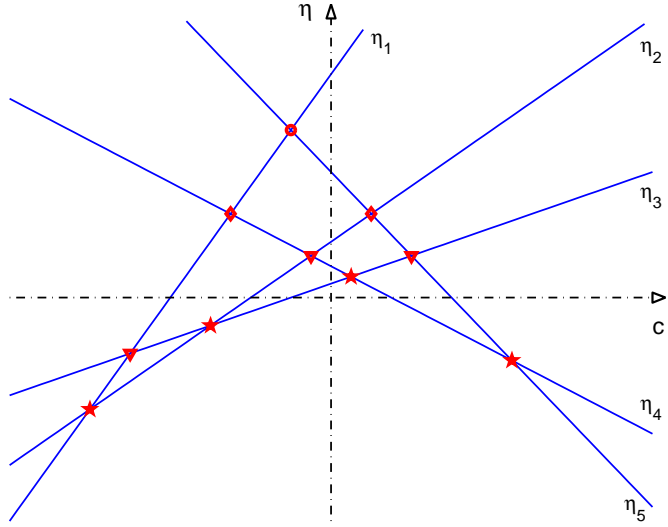}}
\caption{The functions $\eta_j(c)=k_j(k_j-c)$ for
$(k_1,k_2,k_3,k_4,k_5)=(-2,-1,-\frac12,\frac34,\frac32)$.
The levels of intersection 0, 1, 2, 3 are respectively denoted
by circles, diamonds, triangles and stars.}
\label{f:eta}
\end{figure}

Let~us now define a local coordinate frame $(\xi,y)$ in order
to study the asymptotics for large $|y|$ with
\[
x= c\,y + \xi\,.
\]
Then the phase functions $\theta_i$ in $f$ of~\eqref{e:ftoda} become
\[
\theta_i= -k_i\xi + \eta_i(c)\,y + \theta_i^0\,, \quad{\rm for}\quad
i=1,\dots, N\,,
\]
with
\[
\eta_i(c):=k_i(k_i-c)\,.
\]
Without loss of generality, we assume the ordering for the
parameters $\{k_i\,|~i=1,\dots,N\}$,
\[
k_1<k_2<\cdots<k_N\,.
\]
Then one can easily show that the lines $\eta=\eta_i(c)$ are in general
position; that is, each line $\eta=\eta_i(c)$ intersects with all other
lines at $N-1$ distinct points in the $c$-$\eta$ plane;
in other words, only {\it two} lines meets at each intersection point.
Figure~\ref{f:eta} shows a specific example, corresponding to the
values $(k_1,k_2,k_3,k_4,k_5)=(-2,-1,-\frac12,\frac34,\frac32)$.

Now the purpose is to find the dominant exponential terms in the
$\tau_{\Np}$-function~\eqref{e:tauplus} for $y\to \pm\infty$ as a function of
the velocity $c$. First note that if only one exponential is dominant,
then
$w_1=-\partial_x \log\tau_{\Np}$ is just a constant,
and therefore the solution $u=-2\partial_x w_1$ is zero.
Then, nontrivial contributions to $u$
arise when one can find
{\it two} exponential terms which dominate over the others.
Note that because the intersections of the $\eta_i$'s are always pairwise,
three or more terms cannot make a dominant balance for large $|y|$.
In the case of $(\Nm,1)$-soliton solutions, it is easy to see that
at each $c$ the dominant exponential term for $y\to\infty$ is provided
by only $\eta_1$ and/or $\eta_N$, and therefore there is only one shock
($\Np=1$) moving with velocity~$c_{1,N}=k_1+k_N$ corresponding to
the intersection point of $\eta_1$ and $\eta_N$ (see figure \ref{f:eta}).
On the other hand, as $y\to-\infty$, each term $\eta_j$ can become
dominant for some $c$,
and at each intersection point $\eta_j=\eta_{j+1}$
the two exponential terms corresponding to $\eta_j$ and $\eta_{j+1}$
give a dominant balance;
therefore there are $\Nm=N-1$ shocks moving with
velocities~$c_{j,j+1}=k_j+k_{j+1}$, for $j=1,\dots,N-1$
(see again figure~\ref{f:eta}).

In the general case, $\Np\ne1$, the $\tau_\Np$-function
in~\eqref{e:tauplus}
involves exponential terms having combinations of phases, and
two exponential terms that make a dominant balance
can be found as follows:
Let~us first define the {\it level of intersection} of $\eta_i(c)$.

\begin{definition}
Let $\eta_i(c)$ and $\eta_j(c)$ intersect at the value $c=c_{i,j}=k_i+k_j$,
i.e., $\eta_i(c_{i,j})=\eta_j(c_{i,j})$.
The level of intersection, denoted by $\sigma_{i,j}$, is defined
as the number of other $\eta_l$'s that at $c=c_{i,j}$ are larger
than $\eta_i(c_{i,j})=\eta_j(c_{i,j})$.
That is,
\[
\sigma_{i,j}:=\big|\{\eta_l\,|~\eta_l(c_{i,j})>\eta_i(c_{i,j})=\eta_j(c_{i,j})\}\big|\,.
\]
We also define $I(n)$ as the set of pairs $(\eta_i,\eta_j)$ having
the level $\sigma_{i,j}=n$, namely
\[ I(n):=\{(\eta_i,\eta_j)\,|~\sigma_{i,j}=n,~{\rm for}~i<j\,\}.
\]
\end{definition}

\noindent
The level of intersection can take the range
$0\le\sigma_{i,j}\le N-2$.
Then
one can show:

\begin{lemma}
\label{l:intersectingpairs}
The set $I(n)$ is given by
\[
I(n)=\{(\eta_i,\eta_{N-n+i-1})\,|~i=1,\dots,n+1\}\,.
\]
\end{lemma}
\begin{proof}
 From the assumption $k_1<k_2<\cdots<k_N$, we have the following inequality
at $c=c_{i,j}$~ (i.e. $\eta_i=\eta_j$) for $i<j$,
\[
\eta_{i+1},\dots,\eta_{j-1}
<\eta_i=\eta_{j}
<\eta_1,\dots,\eta_{i-1},\,\eta_{j+1},\dots,\eta_N\,.
\]
Then taking $j=N-n-1$ leads to the assertion of the Lemma.
\end{proof} 

\noindent
Note here that the total number of pairs $(\eta_i,\eta_j)$ is
\[
\left(\!\!\begin{array}{cc}N\\2\end{array}\!\!\right)=\half N(N-1)=
             \sum_{n=0}^{N-2}|I(n)|\,.
\]
We illustrate these definitions in figure~\ref{f:eta}, where
the sets $I(n)$ for the level of intersection $n=0, 1, 2, 3$,
which are respectively marked by circles, diamonds, triangles and stars,
are given by
\[
\left\{
\begin{array}{llll}
I(0)=\{(\eta_1,\eta_5)\},\\[0.4ex]
I(1)=\{(\eta_1,\eta_4),~(\eta_2,\eta_5)\},\\[0.4ex]
I(2)=\{(\eta_1,\eta_3),~(\eta_2,\eta_4),~(\eta_3,\eta_5)\},\\[0.4ex]
I(3)=\{(\eta_1,\eta_2),~(\eta_2,\eta_3),~(\eta_3,\eta_4),~(\eta_4,\eta_5)\}.
\end{array}\right.
\]
For the case of $(\Nm,\Np)$-solitons, the following formulae are useful:
\begin{eqnarray*}
\left\{
\begin{array}{ll}
I(\Nm-1)= \{(\eta_i,\eta_{\Np+i})\,|~i=1,\dots,\Nm\}\,,
\\[0.4ex]
I(\Np-1)= \{(\eta_i,\eta_{\Nm+i})\,|~i=1,\dots,\Np\}\,.
\end{array}\right.
\end{eqnarray*}
\noindent
Here recall that $\Np+\Nm=N$.
These formulae indicate that, for each intersecting pair $(\eta_i,\eta_j)$
with the level $\Nm-1$ ($\Np-1$),
there are $\Np-1$ terms $\eta_l$'s which are smaller (larger)
than~$\eta_i=\eta_j$.
Then the sum of those $\Np-1$ terms with either $\eta_i$ or $\eta_j$
provides two dominant exponents in the $\tau_{\Np}$-function
for $y\to -\infty$ $(y\to\infty)$ (see more detail in the proof of
Theorem~\ref{t:asymptotic}).
Note also that $|I(N_\pm-1)|=N_\mp$.
Now we can state our main theorem:

\begin{theorem}
\label{t:asymptotic}
Let\, $w_1$\, be a function defined by
\[
w_1= - \partialderiv{ }x \log \tau_\Np\,,
\]
with $\tau_\Np$ given by~\eqref{e:tauplus}.
Then\, $w_1$\, has the following asymptotics for $y\to\pm\infty:$
\renewcommand\labelenumi{{\rm(\roman{enumi})}}
\begin{enumerate}
\itemsep 0pt
\parsep 0pt
\item
For\, $y\to-\infty$\, and\, $x=c_{i,\Np+i}\,y+\xi$ ~for $i=1,\dots,\Nm$\,,
\[
\kern-\leftmargini
w_1~\longrightarrow~\left\{\begin{array}{ll}
K_i(-,-):= \sum_{j=i+1}^{\Np+i} k_j &\quad\mathrm{as}\quad \xi\to-\infty\,,\\
K_i(+,-):= \sum_{j=i}^{\Np+i-1} k_j &\quad\mathrm{as}\quad \xi\to\infty\,.
\end{array}\right.
\]
\item
For\, $y\to\infty$\, and\, $x=c_{i,\Nm+i}\,y+\xi$~ for $i=1,\dots,\Np$\,,
\[
\kern-\leftmargini
w_1~\longrightarrow~\left\{\begin{array}{ll}
K_i(-,+):=  \sum_{j=1}^{i-1} k_j
             + \sum_{j=1}^{\Np-i+1} k_{N-j+1} &\quad\mathrm{as}\quad
\xi\to-\infty\,,\\
K_i(+,+):= \sum_{j=1}^i k_j
             + \sum_{j=1}^{\Np-1} k_{N-j+i} &\quad\mathrm{as}\quad
\xi\to\infty\,.
\end{array}\right.
\]
\end{enumerate}
where\, $c_{i,j}=k_i+k_j$.
\end{theorem}

\begin{proof}
First note that at the point $\eta_i=\eta_{\Np+i}$, i.e.,
$(\eta_i,\eta_{\Np+i})\in I(\Nm-1)$, from Lemma~\ref{l:intersectingpairs}
we have the inequality,
\[
\underbrace{\phantom{\bigg|}\kern-0.2em
\eta_{i+1},\eta_{i+2},\dots,\eta_{i+\Np-1}}_{\Np-1}<\eta_i=\eta_{\Np+i}\,.
\]
This implies that, for $c=k_i+k_{\Np+i}$,
the following two exponential terms in the
$\tau_\Np$-function in Lemma~\ref{l:tauN},
\[
\exp\bigg(\sum_{j=i}^{\Np+i-1} \theta_j\bigg)\,, 
\qquad%\quad\mathrm{and}\quad
\exp\bigg(\sum_{j=i+1}^{\Np+i} \theta_j\bigg) \,,
\]
provide the dominant terms for $y\to-\infty$.
Note that the condition $\eta_i=\eta_{\Np+i}$ leads to
$c=c_{i,\Np+i}=k_i+k_{\Np+i}$.
Thus the function $w_1$ can be approximated by the following form along
$x=c_{i,\Np+i}\,y+\xi$ for $y\to -\infty$:
\[
\kern-0.5em
\begin{array}{llll}
w_1&\sim&\displaystyle -\partialderiv{}\xi\log\big(
\Delta_i(+,-)e^{-K_i(+,-)\xi}+\Delta_i(-,-)e^{-K_i(-,-)\xi} \big)
\\[1.6ex]
&=&\displaystyle{
\frac{K_i(+,-)\Delta_i(+,-)e^{-K_i(+,-)\xi}+K_i(-,-)\Delta_i(-,-)e^{-K_i(-,-)\xi}}
             {\Delta_i(+,-)e^{-K_i(+,-)\xi}+\Delta_i(-,-)e^{-K_i(-,-)\xi}}\,,}
\\[1.6ex]
&=& \displaystyle{
\frac{K_i(+,-)\Delta_i(+,-)e^{(k_{\Np+i}-k_i)\xi}+K_i(-,-)\Delta_i(-,-)}
             {\Delta_i(+,-)e^{ (k_{\Np+i}-k_i)\xi}+\Delta_i(-,-)}\,,}
\end{array}
\]
where
\begin{eqnarray*}
\Delta_i(+,-)= \Delta(i,\dots,\Np+i-1)\,\,
             \exp\bigg(\sum_{j=i}^{\Np+i-1} \theta_j^0\bigg)
\\
\Delta_i(-,-)= \Delta(i+1,\dots,\Np+i)\,\,
             \exp\bigg(\sum_{j=i+1}^{\Np+i} \theta_j^0\bigg)\,.
\end{eqnarray*}
Now, from $k_i<k_{\Np+i}$~it is obvious that $w_1$ has the desired asymptotics
as $\xi\to\pm\infty$ for $y\to-\infty$.

Similarly,
for the case of $(\eta_i,\eta_{\Nm+i})\in I(\Np-1)$ we have the inequality
\[
\eta_i=\eta_{\Nm+i}<
\underbrace{\phantom{\bigg|}\kern-0.2em
\eta_1,\eta_2,\dots,\eta_{i-1},\eta_{\Nm+i+1}\dots,\eta_N}_{\Np-1}\,.
\]
Then the dominant terms in the $\tau_\Np$-function on $x=c_{i,\Nm+i}y+\xi$ for
$y\to\infty$ are given by the exponential terms
\[
\exp\bigg(\sum_{j=1}^i \,\theta_j ~
             + \sum_{j=1}^{\Np-i} \,\theta_{N-j+1}\bigg)\,,
\qquad %\quad {\rm and} \quad
\exp\bigg(\sum_{j=1}^{i-1} \,\theta_j ~
             + \sum_{j=1}^{\Np-i+1} \,\theta_{N-j+1}\bigg)\,.
\]
Then, following the previous argument, we obtain the desired asymptotics
as $\xi\to\pm \infty$ for $y\to\infty$.

For other values of $c$, that is for $c\ne c_{i,\Np+i}$
and $c\ne c_{i,\Nm+i}$, just one exponential term becomes dominant,
and thus $w_1$ approaches a constant as $|y|\to\infty$.
This completes the proof.
\break{$\phantom.$}
\end{proof}

\begin{figure}[t!]
\medskip
\centerline{\epsfxsize0.925\textwidth\epsfbox{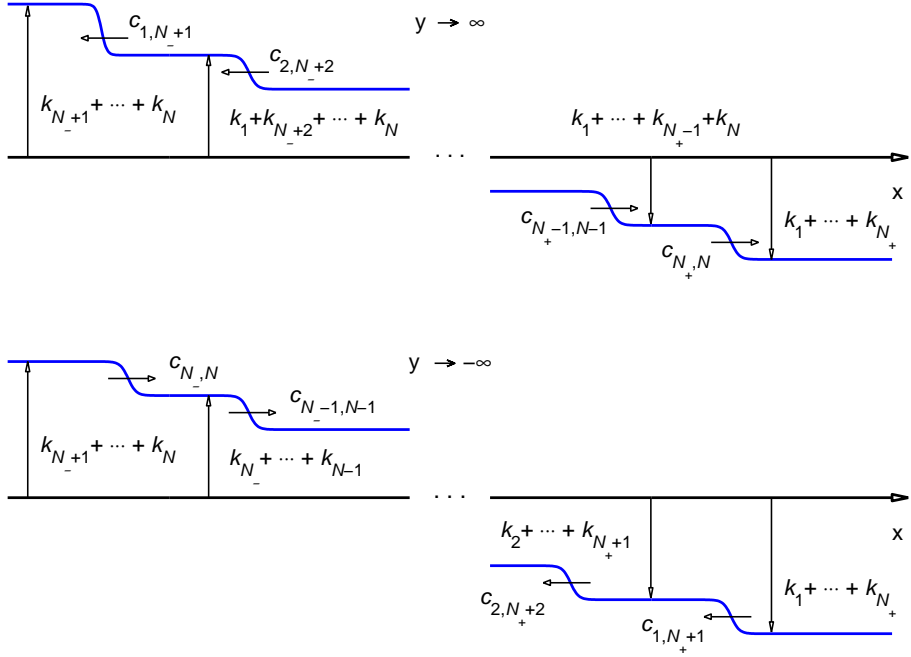}}
\caption{Asymptotic behavior of the function $w_1$ with
$k_1<k_2<\cdots\le0<\cdots<k_{N-1}<k_N$.
As $y\to-\infty$ there are $\Nm$ jumps, moving with velocities
$c_{j,\Np+j}$ ($j=1,\dots,\Nm$).
As $y\to\infty$ there are $\Np$ jumps, moving with velocities
$c_{i,\Nm+i}$ ($i=1,\dots,\Np$).}
\label{f:asymptotic}
\end{figure}

Theorem~\ref{t:asymptotic} can be summarized in figure~\ref{f:asymptotic}:
As $y\to-\infty$, the function $w_1$ has $\Nm$ jumps, moving with velocities
$c_{j,\Np+j}$ for $j=1,\dots,\Nm$;
as $y\to\infty$, $w_1$ has $\Np$ jumps, moving with velocities
$c_{i,\Nm+i}$ for $i=1,\dots,\Np$.
Each jump represents a line soliton of the $u$-solution,
and therefore the whole solution represents an
$(\Nm,\Np)$-soliton.
Each velocity of the asymptotic line solitons in the $(\Nm,\Np)$-soliton
is determined from the $c$-$\eta$ graph of the levels of intersections
(see figure~\ref{f:eta}).
For example, in the case of $(1,4)$-soliton in figure~\ref{f:eta},
one incoming soliton has velocity $c_{1,4+1}=c_{1,5}$, corresponding
to the set $I(0)$, and four outgoing solitons have the velocities $c_{i,1+i}$
for $i=1,\dots,4$, corresponding to $I(3)$.
Note that, given a set of $N$ phases (as determined by the parameters
$k_i$ for $i=1,\dots,N$), the same graph can be used for
any $(\Nm,\Np)$-soliton with $\Nm+\Np=N$.
In particular, if $N=2M$, we have $\Np=\Nm=M$, and Theorem~\ref{t:asymptotic}
implies that the velocities of the $M$ incoming solitons are equal to those
of the $M$ outgoing solitons.
However, we show in the next section that these (resonant) $M$-soliton
solutions are different from the ordinary (nonresonant) multi-soliton
solutions of the KP~equation.

We remark that
Theorem~\ref{t:asymptotic} determines the complete structure of asymptotic
patterns of the solutions $u(x,y,t)$ given by~\eqref{e:kpsoln} for the
Toda lattice equation. In the case of the ordinary multi-soliton solution
of the KP equation, the tau-function~\eqref{e:tau} does not
contain all the possible combinations of phases,
and therefore the theorem should be modified. However, the key idea for
the asymptotic analysis of using the levels of intersection is still
applicable. In fact, one can find from the same argument that the
asymptotic velocities for the ordinary $M$-solitons are given by
$c_{2i-1,2i}=k_{2i-1}+k_{2i}$ where the $\tau_M$-function is the
Wronskian
(\ref{e:tau}) with $f_{i}=e^{\theta_{2i-1}}+e^{\theta_{2i}}$ for
$i=1,\dots,M$ and $k_1<k_2<\cdots<k_{2M}$.
Note that the velocities are different from those of the resonant
$M$-soliton solution for the Toda case.

Finally, it should be noted that the asymptotic values
$w_{1,j}:=-\partial_x\log\tau_j$ as $\xi\to\pm\infty$
show the sorting property of the Toda lattice equation;
that is,
for $j=1,\dots,N$,
\[
b_j=\partialderiv{ }x\log\frac{\tau_j}{\tau_{j-1}}
            = w_{1,j-1}-w_{1,j} ~\longrightarrow~
\left\{
\begin{array}{ll}
-k_j        &\quad\mathrm{as}\quad \xi\to\infty\,,\\
-k_{N-j+1}  &\quad\mathrm{as}\quad \xi\to-\infty\,.
\end{array}\right.
\]
Also, one can easily show that $a_j\to0$ as $|\xi|\to\infty$,
which implies the sorting behavior, i.e.,
\[
L~\longrightarrow~\left\{
\begin{array}{ll}
\diag(-k_1,-k_2,\dots,-k_N)     &\quad\mathrm{as}\quad \xi\to\infty\,,\\
\diag(-k_N,-k_{N-1},\dots,-k_1) &\quad\mathrm{as}\quad \xi\to-\infty\,.
\end{array}\right.
\]
Recall here that the set $\{\lambda=-k_i\,|~i=1,\dots,N\}$ contains the
eigenvalues of the Lax matrix $L$, with $\lambda_1>\cdots>\lambda_N$
as mentioned in~\eqref{e:todasorting}.

%%%%%%%%%%%%%%%%%%%%%%%%%%%%%%%%%%%%%%%%%%%%%%%%%%%%%%%%%%%%%%%%%%%%%%%%%%%%%%%

\section{Intermediate patterns of soliton interactions}

\begin{figure}[t!]
\centerline{\epsfxsize0.35\textwidth\epsfbox{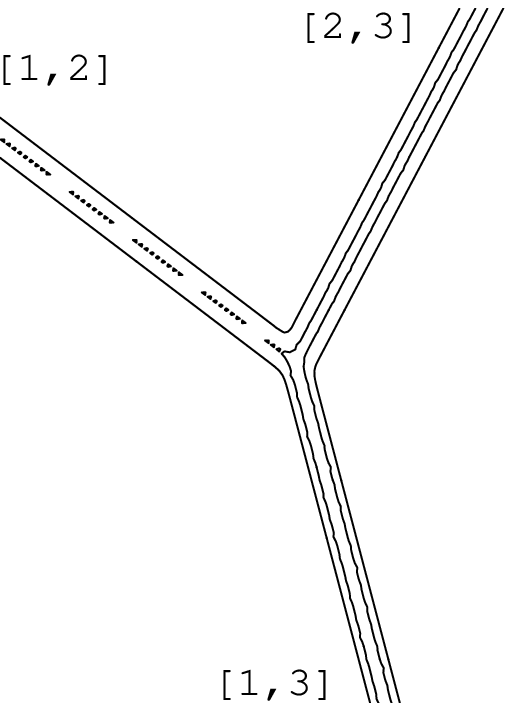}\quad%
\epsfxsize0.495\textwidth\epsfbox{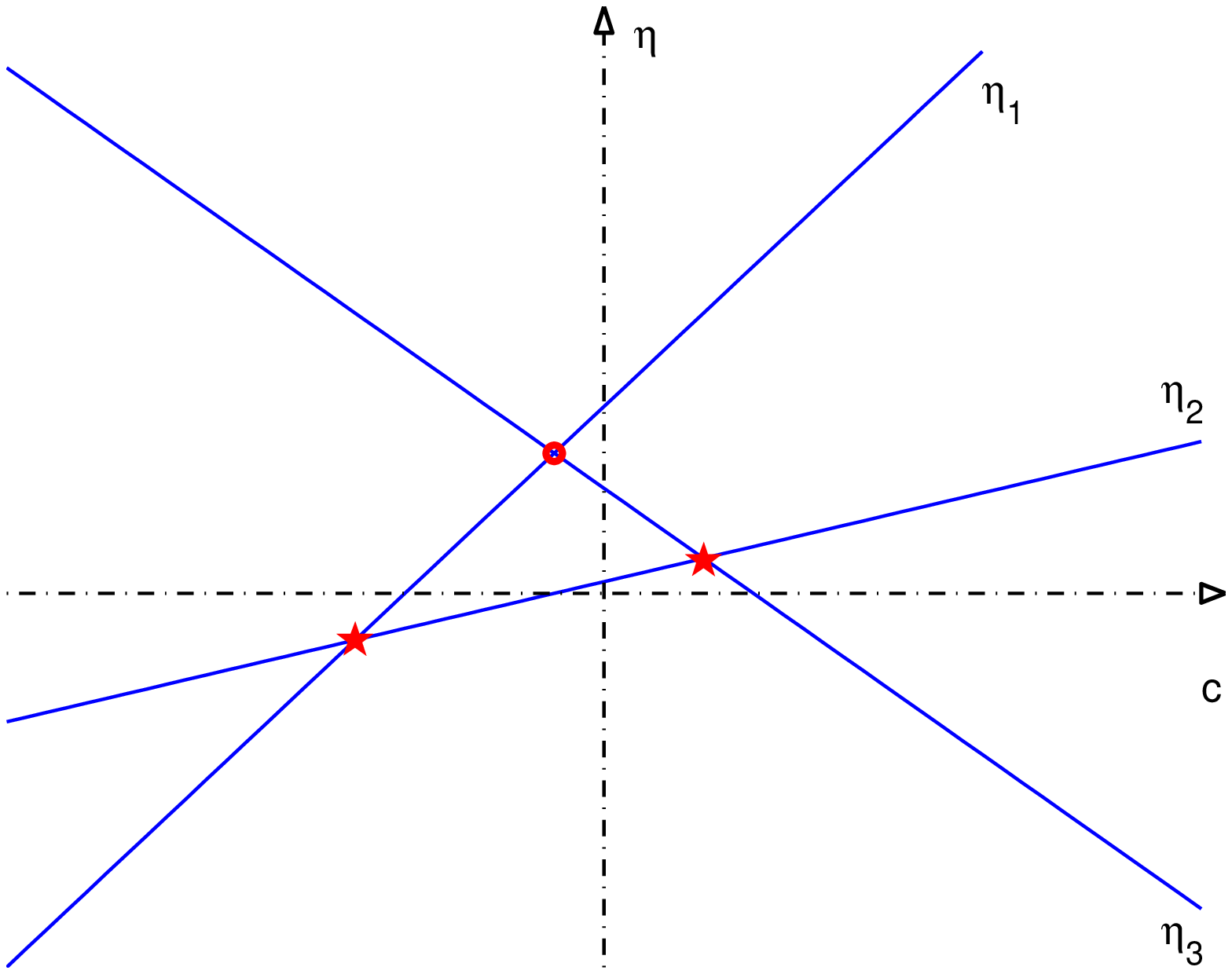}}
\caption{The Y-shape graph (left) illustrating a fundamental
resonance with $(k_1,k_2,k_3)=(-1,-\frac14,\frac34)$ and the
corresponding functions $\eta_i(c)$, $i=1,2,3$ (right).
The graph to the left represents contour lines of $u(x,y,t)$.
The circle at the level set $I(0)$ corresponds to the incoming soliton,
and the stars at $I(1)$ correspond to the outgoing solitons.}
\label{f:miles}
\end{figure}

In this section we describe the intermediate patterns of the resonant
solitons in the $x$-$y$~plane.
The key idea is to consider the pattern as a collection of fundamental
resonances.
The fundamental resonance consists of three parameters:
$\{k_1,k_2,k_3\}$,
that is, the case of $N=3$ with $|\Nm-\Np|=1$.
Without loss of generality, let us take $\Nm=1$ and $\Np=2$,
i.e., a (1,2)-soliton.
(The case of a (2,1)-soliton is obtained from the symmetry
$(x,y,t)\to (-x,-y,-t)$ of the KP equation, i.e., from the
duality of the determinants, $\tau_1$ and $\tau_2$ for $N=3$.)\,\
Then, with $k_1<k_2<k_3$, the pattern of the fundamental resonance
is a Y-shape graph as shown in figure~\ref{f:miles}.
Here and in the following we denote with $[i,j]$
the {\it asymptotic} line soliton with $c=c_{i,j}=k_i+k_j$.
Notice that $I(\Nm-1)=I(0)=\{(\eta_1,\eta_3)\}$
and $I(\Np-1)=I(1)=\{(\eta_1,\eta_2),(\eta_2,\eta_3)\}$.

One should note that at the vertex of the Y-shape graph
each index appears exactly twice as the result of resonance,
and in figure~\ref{f:miles}b those vertices form a triangle,
which we refer to as a ``resonant triangle''.
The resonant triangle is equivalent to the
resonance condition for the wavenumber vectors in~\eqref{resonant}.
Since the vertex of the Y-shape graph consists of three line solitons,
$\theta_i=\theta_j,~1\le i<j\le 3$, the location of the vertex is
obtained from the solution of the equations
$\theta_1=\theta_2=\theta_3$, i.e.,
\[
\left(\!\begin{array}{cc}
k_1-k_2 &-(k_1^2-k_2^2)\\
k_1-k_3 &-(k_1^2-k_3^2)
\end{array}\!\right)
\left(\!\!\begin{array}{c}x\\y\end{array}\!\!\right)
=
\left(\!\!\begin{array}{c}
\theta_1^0-\theta_2^0-(k_1^3-k_2^3)t\\
\theta_1^0-\theta_3^0-(k_1^3-k_3^3)t
\end{array}\!\!\right)\,.
\]
Note here that the coefficient matrix is nonsingular for $k_1<k_2<k_3$, and
the location $(x,y)$ is uniquely determined by a function of $t$.
This implies that there always exists a Y-shape graph if there are three
line solitons satisfying the resonance conditions (\ref{resonant}).
Since the $\tau_{\Np}$-function (\ref{e:tauplus}) contains
all possible combinations of $\Np$ phases, all the vertices in the
graph form Y-shape intersections as a result of dominant balance of
three exponential terms in the $\tau_{\Np}$ at each vertex.
One should also note that a vertex with 4 or more line solitons is not generic:
A vertex with $m$ {\it distinct} line solitons is obtained from the system
of  $m$ equations, $\{\theta_{i_k}=\theta_{j_{k}}~|~i_k\ne j_k,~k=1,\dots,m\}$,
in which at least $m-1$ equations are linearly independent.
Then for $m\ge 4$,
this system in $(x,y)$ is overdetermined, so that the solution exists
only for specific choices of $\theta_i^0$ for fixed values of~$t$.
In the cases of both ordinary and resonant 2-soliton solutions, the two
pairs of solitons as $y\to\pm\infty$ are the same, and therefore there are
only two independent equations.
Also, as mentioned before, the ordinary 2-soliton solution needs a balance
of four exponential terms to realize an X-shape vertex.
However, this balance cannot be dominant over a balance of
three terms with the $\tau_{\Np}$-function given by~\eqref{e:tauplus}.
In what follows, we show that the X-shape vertex of an ordinary
2-soliton solution is blown up into a hole with four Y-shape vertices
for the resonant 2-soliton solution.

\begin{figure}[t!]
\medskip
\centerline{\epsfxsize0.425\textwidth\epsfbox{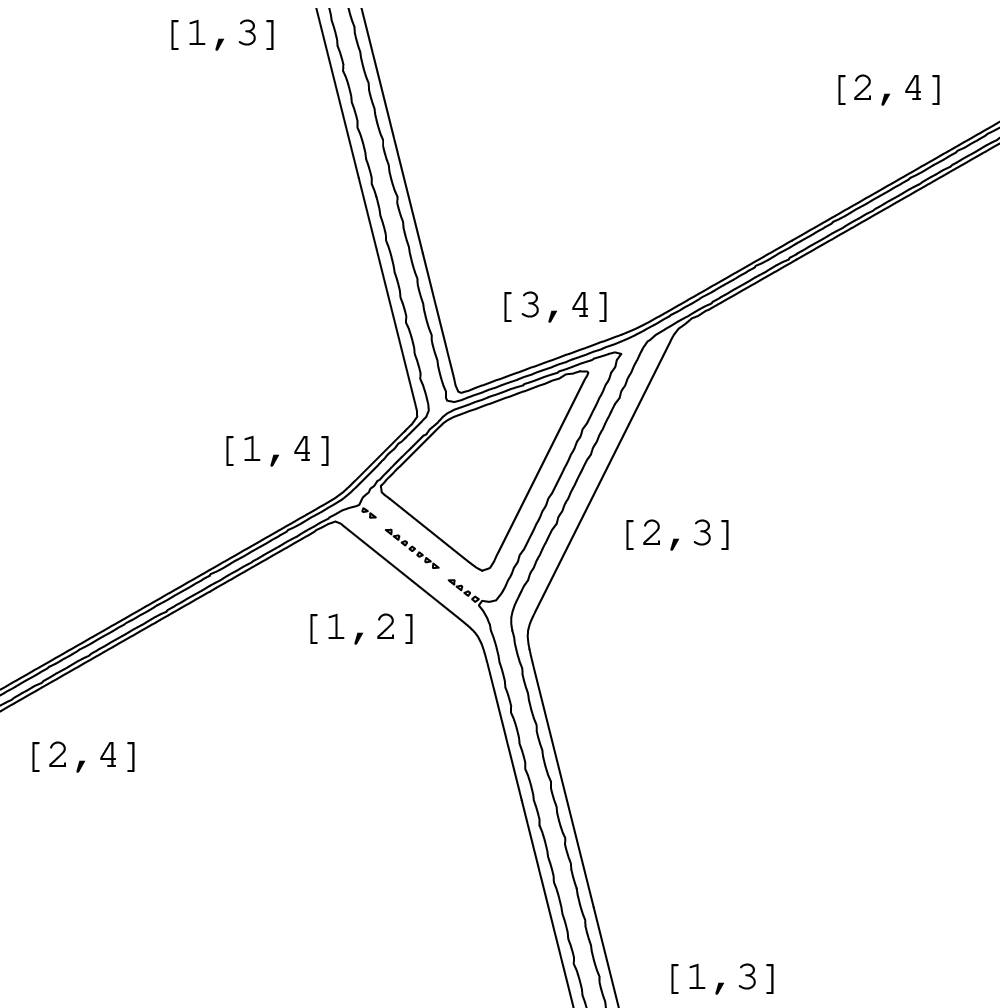}\quad%
\raise2ex\hbox{\epsfxsize0.525\textwidth\epsfbox{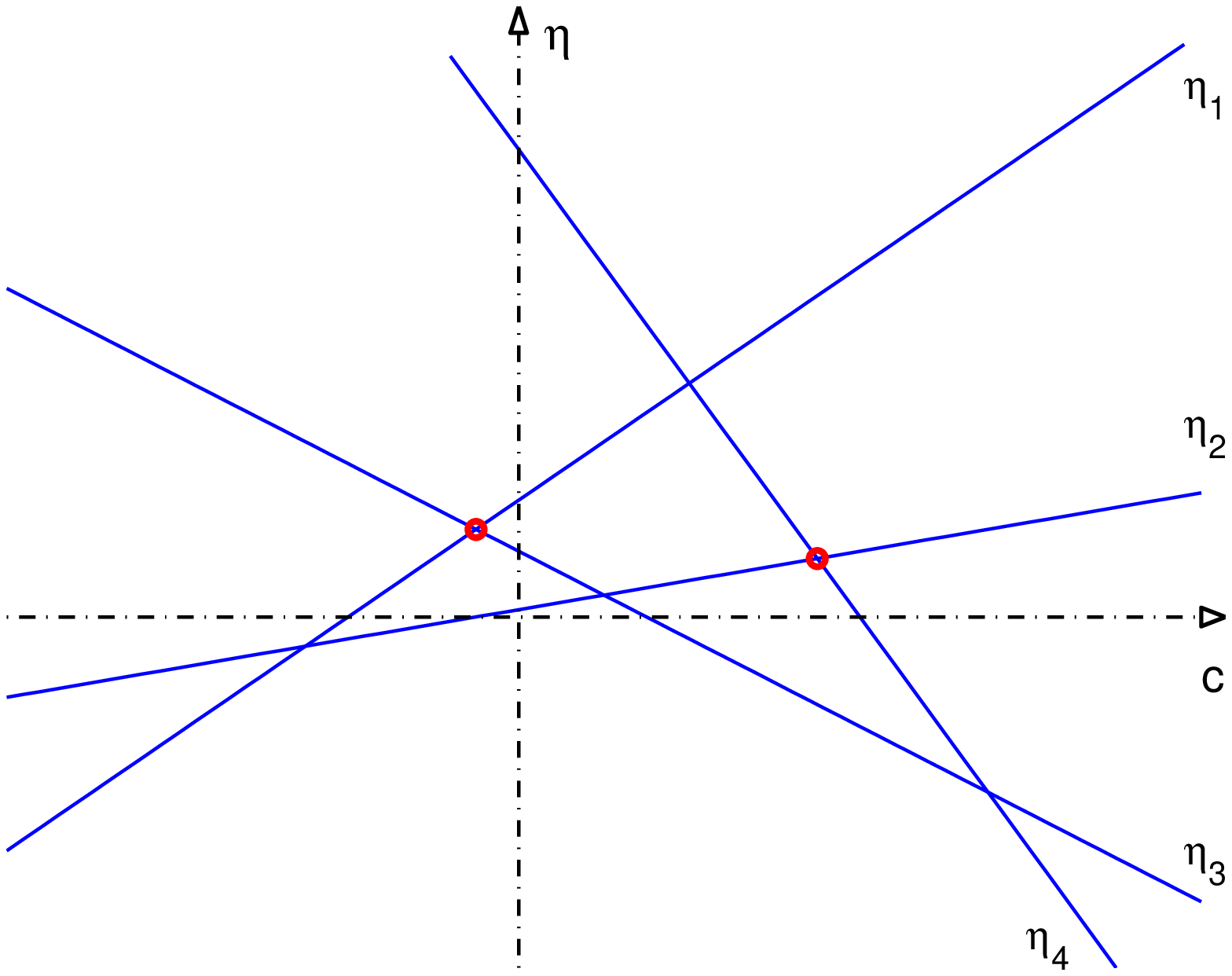}}}
\caption{A resonant 2-soliton solution $u(x,y,t)$ (left) with
$(k_1,k_2,k_3,k_4)=(-1,-\frac14,\frac34,2)$ and the corresponding
functions $\eta_i(c)$ (right). Both incoming and outgoing solitons
correspond to the interstections marked by circles at the level
set $I(1)$.}
\label{f:2-2resonance}
\end{figure}

We now consider the case with $\Nm=2$ and $\Np=2$, which describes the
resonant 2-soliton solution.
We can start with the graph in figure~\ref{f:miles} having
$k_1<k_2<k_3$.
Then we add $k_4$ with $k_3<k_4$.  From
Theorem~\ref{t:asymptotic} we find that both asymptotic solutions for
$y\to\pm\infty$ consist of the solitons with
[1,3] and [2,4].
With $k_1<k_2<k_3<k_4$, the velocity $c_{2,4}$ of the additional
soliton [2,4] as $y\to-\infty$ satisfies $c_{2,4}>c_{2,3}>c_{1,2}$.
For sufficiently large negative values of $t$,
the [2,4]~soliton starts in the left side of the [1,3]~soliton
and first intersects with the [1,2]~soliton;
then the resonance condition determines that the [1,2] and [2,4]~solitons
merge and make a new outgoing soliton~[1,4].
Since the $\Np$ solitons consist of [1,3] and~[2,4],
this [1,4]~soliton first branches to [1,3] and~[3,4].
Then the intermediate [3,4]~soliton now intersects with the
[2,3] soliton to form the [2,4]~outgoing soliton.
(Note that $c_{3,4}$ is the largest velocity among these solitons.)
The process forming a resonant 2-soliton is shown in
figure~\ref{f:2-2resonance}. Note here that there are four vertices
in the interaction pattern, which correspond to the four resonant
triangles in the $c$-$\eta$ plane.

One should also note that the [2,4]~soliton cannot intersect directly the
[1,3]~soliton
unless a [1,2]~soliton or a [3,4]~soliton are created as intermediate
solitons.
The graph of this latter case is obtained from figure~\ref{f:2-2resonance}
by letting $(x,y,t)\to (-x,-y,-t)$.
Also note that the ordinary 2-soliton solution with those same parameters
$(k_1,\dots,k_4)$ for $\{f_1,f_2\}$ in~\eqref{e:f2solitons} has different
asymptotic solitons, namely [1,2] and~[3,4], and, because of the missing
exponential terms in the $\tau_2$-function, this ordinary 2-soliton solution
cannot have resonant interactions; that is, no resonant triangle can be
formed with only those exponential terms.
This is also true for any ordinary multi-soliton solutions
of the KP equation.

We can continue the process of adding $n$ new incoming solitons to the
graph in figure~\ref{f:2-2resonance} to get a
$(2+n,2)$-soliton solution.
One can also add $m$ new outgoing solitons to the new graph to obtain
a $(2+n,2+m)$-soliton solution.
This last step can be done by adding $m$ {\it incoming} solutions to
a $(2,2+n)$-soliton solution, which is simply obtained by the $\pi$ rotation
(i.e., $(x,y)\to (-x,-y)$) of the graph of the $(2+n,2)$-solution using the
duality of the determinant.
Then one can show the following:

\begin{proposition}
\label{p:holes}
In the generic situation, the number of holes (bounded regions) in the
graph of the $(\Nm,\Np)$-soliton solution is $(\Nm-1)(\Np-1)$.
\end{proposition}

\begin{figure}[t!]
\medskip
\centerline{\epsfxsize0.825\textwidth\epsfbox{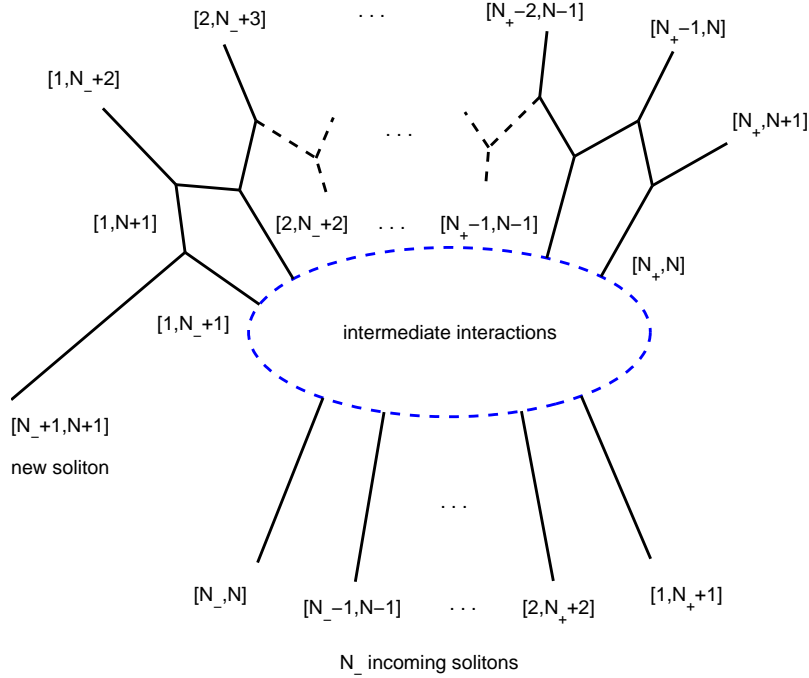}}
\caption{A schematic diagram illustrating the creation of new holes
in the resonant interaction process
for a $(\Nm+1,\Np)$-soliton~solution with $\Nm+\Np=N$. The new soliton
$[\Nm+1,N+1]$ is assumed to have a resonant interaction with the previous
outgoing soliton $[1,\Nm+1]$.}
\label{f:multires}
\end{figure}

\begin{proof}
We use mathematical induction.
The case $\Np=1$ corresponds to the Burgers equation,
and it is immediate to show that the graph of the
(\Nm,1)-soliton solution has a tree shape; that is, no holes
(see also Ref.~\cite{LMP2002v62p91}).
Now suppose that the $(\Nm,\Np)$-soliton has $(\Nm-1)(\Np-1)$ holes.
Add a new phase $\theta_{N+1}$, with $k_{N+1}$ satisfying 
$k_1<\cdots<k_N<k_{N+1}$,
which produces a new, fastest, incoming $[\Nm+1,N+1]$~soliton,
and assume that this solution intersects with the $[1,\Nm+1]$~soliton,
which is the slowest outgoing soliton.
Then the resonant process of those solitons generates a
$[1,N+1]$-soliton as a (2,1)~process, which then intersects
with the new slowest $[1,N+2]$~soliton to generate an intermediate
$[\Nm+2,N+1]$~soliton.
This intermediate soliton interacts with the second slowest outgoing
soliton, the $[2,N+2]$ soliton, to generate
$[2,N+3]$ and $[N+3,N+1]$ solitons, and so on.
This process is illustrated in figure~\ref{f:multires}. From this
figure, it is obvious that there are $\Np-1$
newly created holes; that is, %we have that,
if $(\Nm,\Np)\to(\Nm+1,\Np)$, the number of holes increases as
\[
(\Nm-1)(\Np-1)\to (\Nm-1)(\Np-1) + (\Np-1)=\Nm(\Np-1)\,.
\]
The case of the $(\Nm,\Np+1)$~solution can be analyzed in the same way
using the duality of the determinants.
This completes the proof.
\end{proof}

One can also show the following from Proposition \ref{p:holes}:

\begin{corollary} 
In the generic situation for $\Nm+\Np=N\ge 3$,
the total numbers of intersection points and
intermediate solitons in a $(\Nm,\Np)$-soliton solution are
respectively given by\,\
$2\Nm\Np-N$\,\ and\,\ $3\Nm\Np-2N$.
\end{corollary}

\begin{proof}
By applying mathematical induction on figure \ref{f:multires},
one can easily find that the number of new vertices
(intersection points) is $2\Np-1$ and that of new intermediate
solitons is $3\Np-2$.  This yields the desired results.
\end{proof}

One should compare these numbers with the case of ordinary $M$-soliton
solution, where the total numbers of holes and
intersection points are $\half (M-1)(M-2)$ and $\half M(M-1)$,
respectively.
The resonant process blows up each vertex in an ordinary
$M$-soliton solution to create a hole, so that the total number
of holes in a resonant $M$-soliton solution is given by
\[
\half (M-1)(M-2)+\half M(M-1)=(M-1)^2\,.
\]
Note also that the total number of vertices in a
resonant $M$-soliton is four times of the vertices of an
ordinary $M$-soliton, i.e. each vertex is blown up to make
4 vertices with one hole.

\begin{figure}[t!]
\bigskip
\hbox to\textwidth{\hss\qquad(a)~\quad\qquad~(b)\quad\hss}
\kern-1.2\bigskipamount
\centerline{\epsfxsize0.475\textwidth\epsfbox{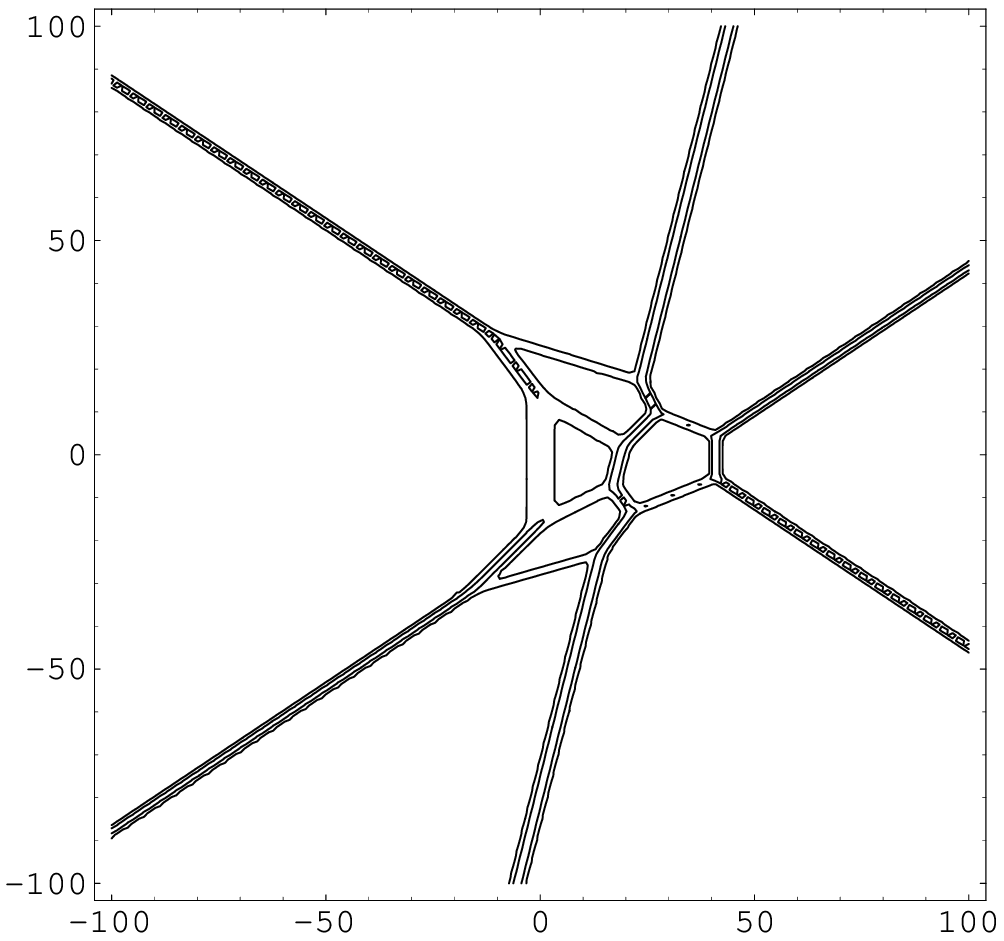}\quad%
\epsfxsize0.475\textwidth\epsfbox{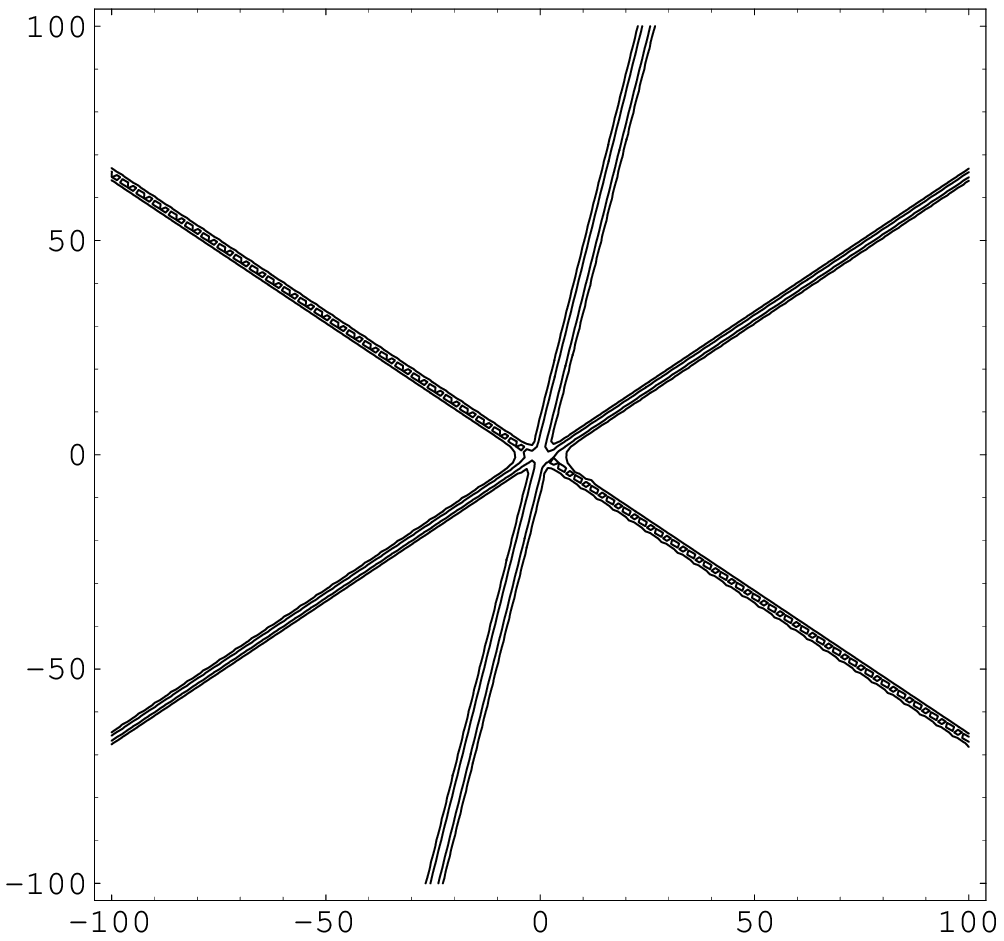}}
\bigskip
\hbox to\textwidth{\hss\qquad(c)~\quad\qquad~(d)\quad\hss}
\kern-1.2\bigskipamount
\centerline{\epsfxsize0.475\textwidth\epsfbox{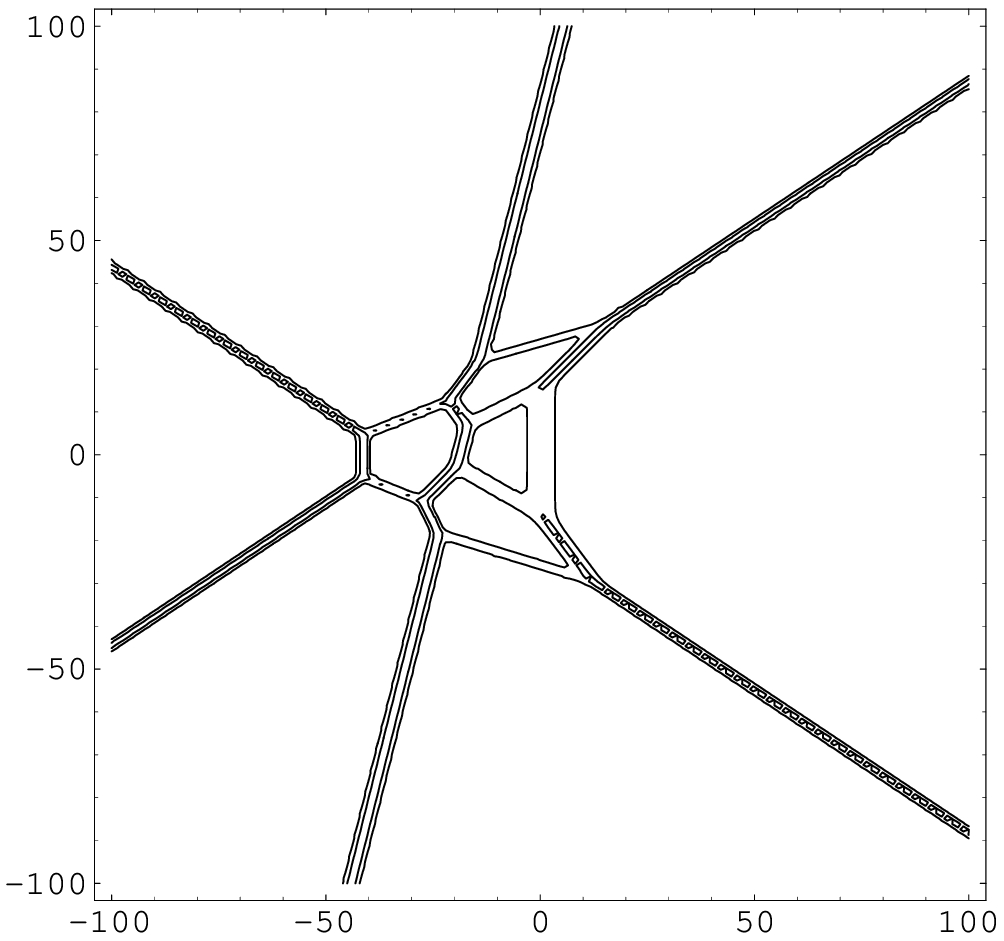}\quad%
\epsfxsize0.475\textwidth\epsfbox{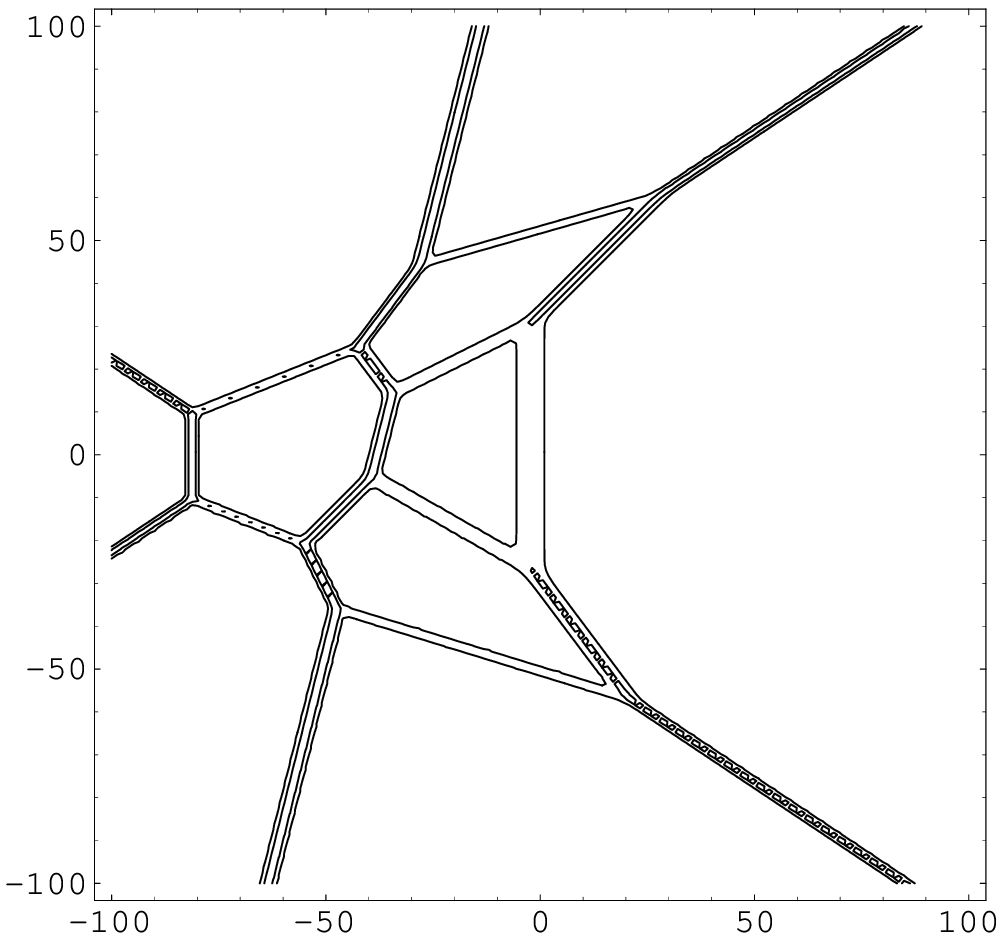}}
\caption{Snapshots illustrating the temporal evolution of a resonant
3-soliton solution $u(x,y,t)$ with
$(k_1,\dots,k_6)=(-\frac52,-\frac54,-\frac12,\frac12,\frac32,\frac52)$
and $\theta_1^0=\cdots=\theta_6^0=0$:
(a)~$t=-10$, (b)~$t=0$, (c)~$t=10$, (d)~$t=20$.
Note the symmetry $(x,y,t)\leftrightarrow (-x,-y,-t)$ in (a) and (c).}
\label{f:res3soliton}
\end{figure}

\begin{figure}[t!]
\bigskip
\hbox to\textwidth{\hss\qquad(a)~\quad\qquad~(b)\quad\hss}
\kern-1.2\bigskipamount
\centerline{\epsfxsize0.475\textwidth\epsfbox{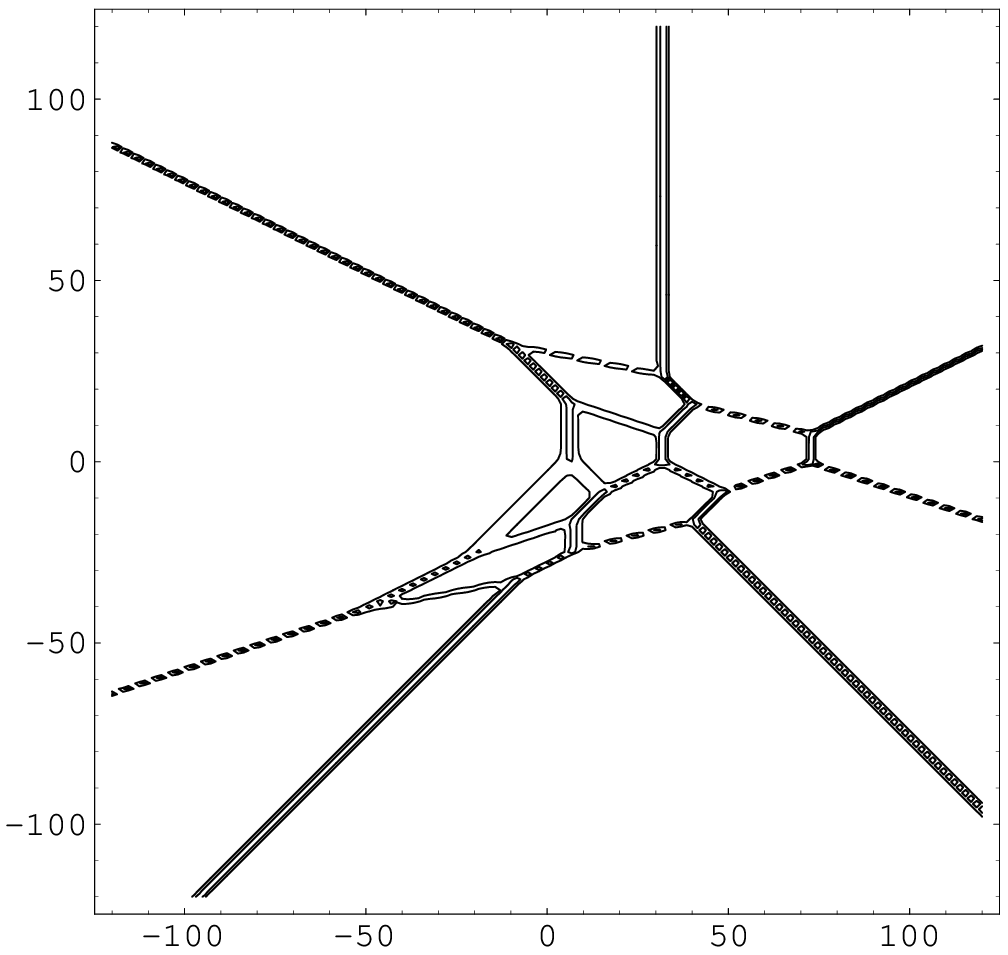}\quad%
\epsfxsize0.475\textwidth\epsfbox{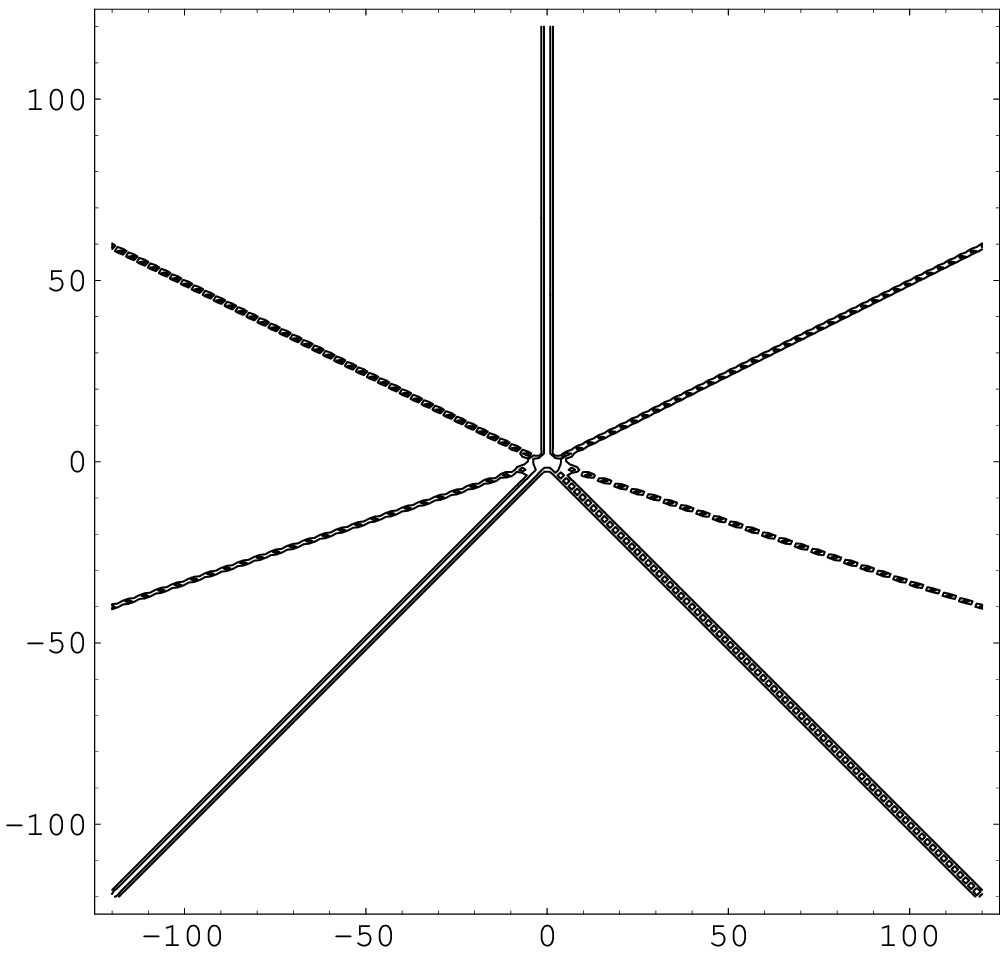}}
\bigskip
\hbox to\textwidth{\hss\qquad(c)~\quad\qquad~(d)\quad\hss}
\kern-1.2\bigskipamount
\centerline{\epsfxsize0.475\textwidth\epsfbox{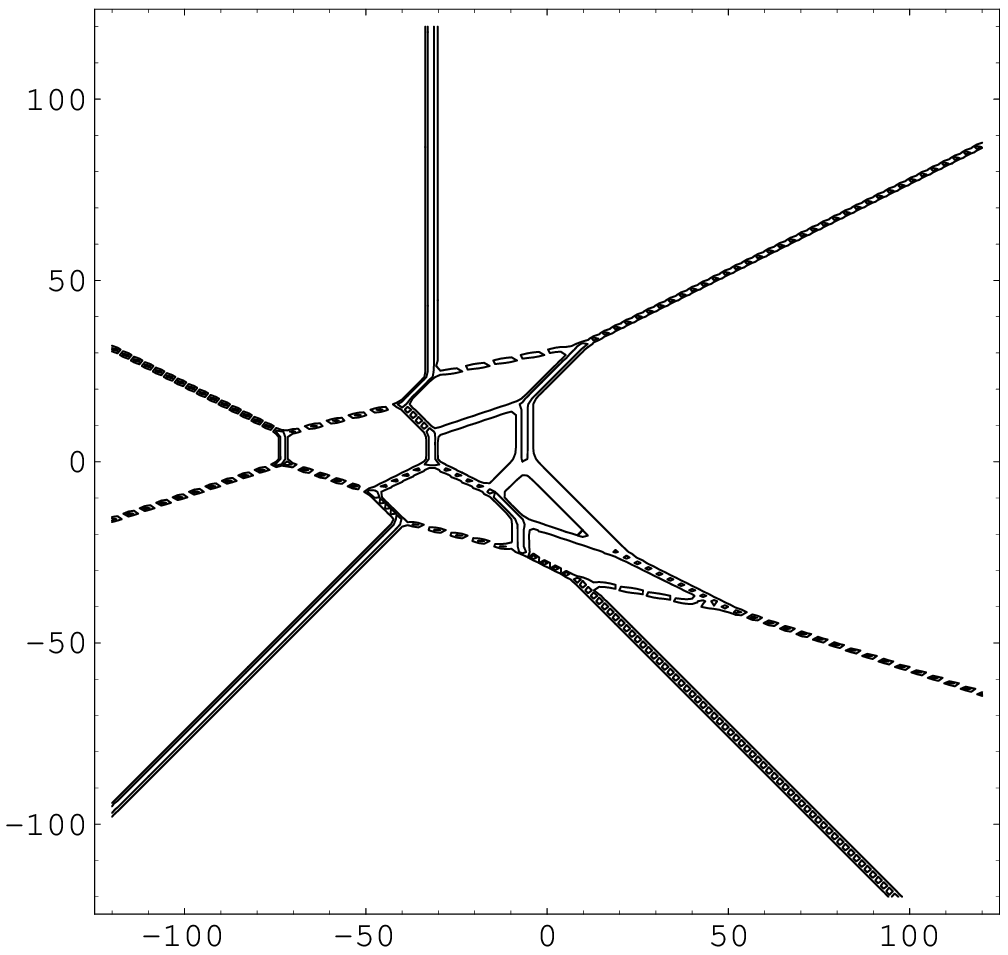}\quad%
\epsfxsize0.475\textwidth\epsfbox{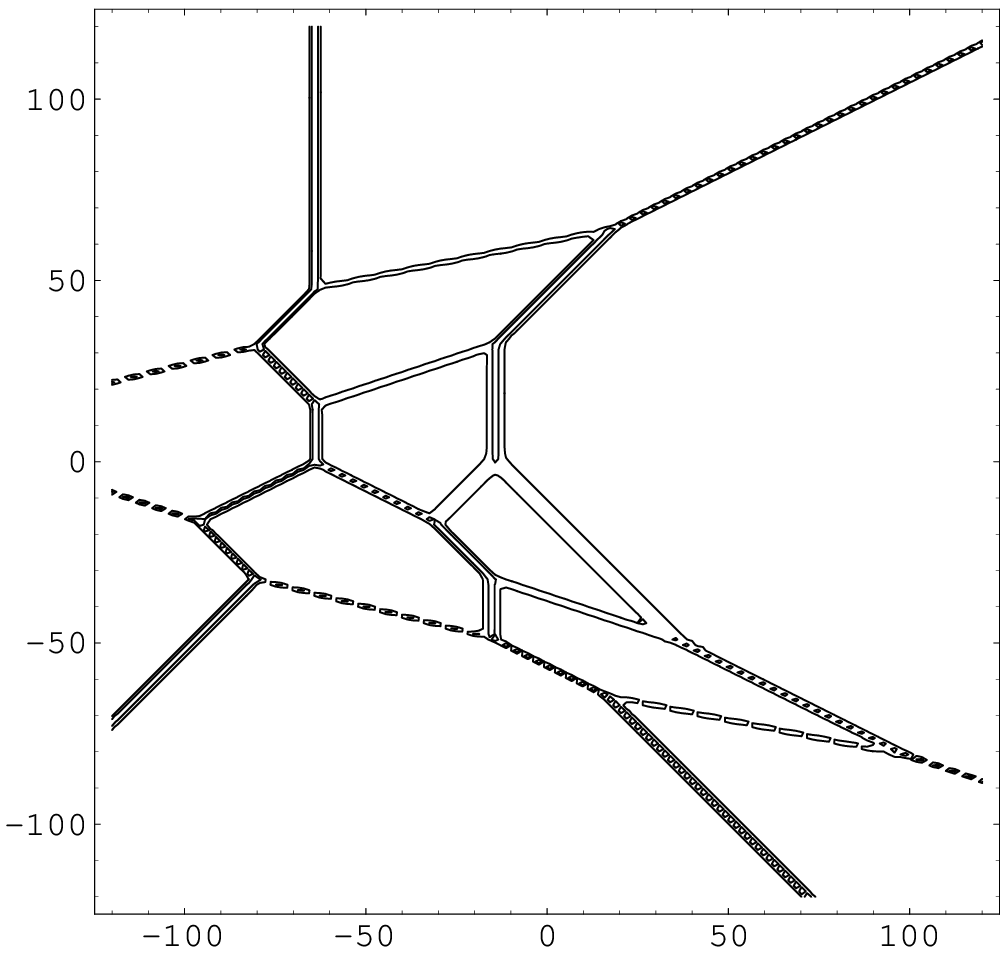}}
\caption{Snapshots illustrating the temporal evolution of a
(4,3)-soliton solution $u(x,y,t)$ with
$(k_1,\dots,k_7)=(-3,-2,-1,0,1,2,3)$ and $\theta_1^0=\cdots=\theta_7^0=0$:
(a)~$t=-8$, (b)~$t=0$, (c)~$t=8$, (d)~$t=16$. Note the symmetry $(x,y,t)
\leftrightarrow (-x,y,-t)$ in (a) and (c).}
\label{f:pattern}
\end{figure}

Figure~\ref{f:res3soliton} shows a few snapshots illustrating the
temporal evolution of a resonant 3-soliton solution with
$(k_1,\dots,k_6)=(-\frac52,-\frac54,-\frac12,\frac12,\frac32,\frac52)$.
This resonant 3-soliton
is similar to the ``spider-web-like" soliton solution found for
the cKP equation (cf. figure 10 in Ref.~\cite{JPhysA2002v35p6893}),
even though the underlying equation is different in those two cases.
As described in this paper, the behavior is determined by the structure of
the tau-function which is just the sum of exponential functions.
The tau-functions of the KP and cKP~equations have the same structure
for those solutions.

Figure~\ref{f:pattern} shows the temporal evolution of
a (4,3)-soliton solution with $(k_1,\dots,k_7)=(-3,-2,-1,0,1,2,3)$.
In both figure~\ref{f:res3soliton} and figure~\ref{f:pattern}, it can be
observed that different intermediate solitons
mediate the interaction process at different times.
Also note that, for some finite values of~$t$,
the number of holes in the solution changes. However,
Proposition~\ref{p:holes} applies in the generic situation,
and the total number of holes remains $(\Nm-1)(\Np-1)$,
namely 4~holes in figure~\ref{f:res3soliton}
and 6~holes in figure~\ref{f:pattern}.
In both figures, we have set
all $\theta_i^0=0$, so that all line solitons merge initially at the origin.
It should be noted that even though several solitons might
merge at the same point for some finite values of $t$, generically the
resonant interactions are always among three solitons, i.e.,
fundamental resonances, as explained in this paper.

Finally, we would like to point out that the KP equation has a
large variety of multi-soliton-type solutions.
Among those solutions, we found that,
since the $\tau_{\Np}$-function of the resonant $(\Nm,\Np)$-soliton
for the Toda lattice hierarchy contains all possible
combinations of phase terms $\{\theta_i\,|~i=1,\dots,N\}$,
the interaction process for these solutions results in a
{\it fully} resonant situation.
On the other hand, the ordinary $M$-soliton solutions display a
{\it nonresonant} case;
that is, resonant triangles representing either (2,1)- or (1,2)-solitons
cannot be formed because of the
missing exponential terms in the tau-function.
One can then find a {\it partially} resonant case consisting
of ordinary multi-soliton interaction with the addition of
some resonant interactions;
one such example is the case having 
$f_1=e^{\theta_1}+e^{\theta_2}+e^{\theta_3}$
and $f_2=\e^{\theta_3}+e^{\theta_4}$ for the $\tau_2$-function~\eqref{e:tau}
where the ordinary 2-soliton interaction coexists with resonant interactions.
We will report the details of the general patterns for multi-soliton-like
solutions for the KP equation in a future communication.

\section*{Acknowledgments}

The work of Y.K.\ was partially supported by NSF grant DMS-0071523;
G.B.\ was partially supported by NSF grant DMS-0101476.

%%%%%%%%%%%%%%%%%%%%%%%%%%%%%%%%%%%%%%%%%%%%%%%%%%%%%%%%%%%%%%%%%%%%%%%%%%%%%%%

\section*{Appendix: The Wronskian solutions of the KP hierarchy}
\appendix
\renewcommand\theequation{A.\arabic{equation}}

In this Appendix we briefly explain how the Wronskian solution~\eqref{e:tau}
is obtained from the Sato theory (see Ref.~\cite{ohta1988} for more details).
The Sato theory is formulated on the basis of a pseudo-differential operator,
\[
{\mathcal L}=\partial +u_2\partial^{-1}+u_3\partial^{-2}+\cdots,
\]
where $\partial$ is a derivation satisfying
$\partial\,\partial^{-1}=\partial^{-1}\,\partial=1$ and
the generalized Leibnitz rule,
\[
\partial^{\nu}(fg)=\sum_{k=0}^{\infty}\bigg(\!\!
\begin{array}{cc}\nu\\k\end{array}\!\!\bigg)\,
\frac{\partial^kf}{\partial x^k}\,\,\partial^{\nu-k}g\,,
\qquad\mathrm{for}\quad\nu\in{\mathbb Z}\,.
\]
(Note that the series terminates if and only if $\nu$ is a positive integer.)
Then the KP hierarchy can be written in the Lax form
\begin{equation}
\label{e:kpl}
\frac{\partial {\mathcal L}}{\partial t_n}=[{\mathcal B}_n,{\mathcal L}],
\quad\mathrm{with}\quad
{\mathcal B}_n:=({\mathcal L}^n)_{\ge 0},
\end{equation}
where $({\mathcal L}^n)_{\ge0}$ represents the polynomial
(differential) part of ${\mathcal L}^n$ in $\partial$. Here the
solution of the KP equation~\eqref{e:kp} is given by $u=2u_2$ with
$t_1=x,~t_2=y$ and $t_3=t$.

Now writing ${\mathcal L}$ in the dressing form,
\[
{\mathcal L}={\mathcal W}{\partial}{\mathcal W}^{-1}, \quad {\rm with}\quad
{\mathcal W}=1+w_1\partial^{-1}+w_2\partial^{-2}+\cdots,
\]
the KP hierarchy becomes
\begin{equation}
\label{e:weq}
\frac{\partial{\mathcal W}}{\partial t_n}={\mathcal B}_n{\mathcal
W}-{\mathcal W}\,\partial^n.
\end{equation}
Using~\eqref{e:kpl}, the variables $u_i$'s can be expressed in terms of
$w_j$'s; for example,
\[
\left\{
\begin{array}{ll}
u_2=-w_{1,x},\\[0.4ex]
u_3=-w_{2,x}+w_1w_{1,x},
\end{array}\right.
\]
and so on.
(Here and in the following,
subscripts $x$ and $t_n$ denote partial differentiation.)\,\
The equations for $w_j$ are, for example,
\[
\left\{
\begin{array}{ll}
w_{1,t_2}=-2w_1w_{1,x}+w_{1,xx}+2w_{2,x},\\[0.4ex]
w_{2,t_2}=-2w_2w_{1,x}+w_{2,xx}+2w_{3,x},
\end{array}\right.
\]
and so on.
Here one can easily show that a finite truncation of ${\mathcal W}$,
given by
\[
{\mathcal W}_M:=1+w_1\partial^{-1}+\cdots +w_M\partial^{-M}\,,
\]
is invariant under the equation~\eqref{e:weq}. For example,
the ${\mathcal W}$-equation with $M=1$ truncation, i.e.
${\mathcal W}_1=1+w_1\partial^{-1}$, is just the Burgers equation,
\begin{equation}
\label{e:burgers}
w_{1,t_2}=-2w_1w_{1,x}+w_{1,xx}.
\end{equation}
For the $M$-truncation, consider the ordinary differential equation
for a function $f$,
\begin{equation}
\label{e:fm}
{\mathcal W}_M\partial^Mf=f^{(M)}+w_1f^{(M-1)}+\cdots+w_Mf=0.
\end{equation}
Let $\{f_j\,|~j=1,\dots,M\}$ be a fundamental set of solutions of
(\ref{e:fm}).
Then the coefficient function $w_1$ is expressed in terms of the
Wronskian for the set of those solutions, i.e.,
\[
w_1=-\frac{\partial}{\partial x}\log \tau_M\,,\qquad {\rm with}\quad
\tau_M={\rm Wr}(f_1,\dots,f_M).
\]
which leads to a solution of the KP equation,
\[
\displaystyle{u=2u_2=-2\frac{\partial}{\partial
x}w_1=2\frac{\partial^2}{\partial x^2}\log \tau_M\,.}
\]
Recall that for $M=1$ this equation gives the well-known Cole-Hopf
transformation between the Burgers equation for $w_1$ and the linear
diffusion equation for $\tau_1=f$. One can also show from~\eqref{e:weq}
that $f$ satisfies the linear partial differential
equations,
\[
\frac{\partial f}{\partial t_n}=\frac{\partial ^nf}{\partial x^n}, \quad
{\rm for}\quad n=1,2,\dots.
\]
Thus the equations for $(w_1,\dots,w_M)$ on the $M$-truncation are
linearizable, and the behavior of the solutions is expected to be
similar to the case of the Burgers equation.
(The $M$-truncated equation is a multi-component extension of
the Burgers equation~\cite{harada1987}.)
This is one of the main motivations of the present study.

%%%%%%%%%%%%%%%%%%%%%%%%%%%%%%%%%%%%%%%%%%%%%%%%%%%%%%%%%%%%%%%%%%%%%%%%%%%%%%%

\end{document}